\documentclass[twocolumn,tighten]{aastex62}
\usepackage{xspace}
\usepackage{amsmath}

\newcommand{\Le}{L$_{\rm \nu,850}$\xspace}
\newcommand{\Led}{L$_{\rm \nu,850,direct}$\xspace}
\newcommand{\Lei}{L$_{\rm \nu,850,inferred}$\xspace}
\newcommand{\Se}{S$_{\rm \nu,850}$\xspace}
\newcommand{\Mdust}{M$_{\rm dust}$\xspace}
\newcommand{\Tdust}{T$_{\textnormal{\scriptsize{dust}}}$\xspace}
\newcommand{\HH}{H$_2$\xspace}
\newcommand{\HI}{H {\sc i}}
\newcommand{\CO}[2]{CO~(#1--#2)\xspace}
\newcommand{\Msun}{M$_{\odot}$\xspace}
\newcommand{\kms}{km~s$^{-1}$\xspace}
\newcommand{\mJybeam}{mJy~beam$^{-1}$\xspace}
\newcommand{\Lprime}[3]{L$^{\prime}_{\textnormal{\scriptsize{#1 (#2--#3)}}}$\xspace}
\newcommand{\Mmol}{M$_{\rm mol}$\xspace}
\newcommand{\Mmole}{M$_{\rm mol,850}$\xspace}
\newcommand{\Mmolt}{M$_{\rm mol,true}$\xspace}
\usepackage{color,comment}

\begin{document}

\correspondingauthor{G. C. Privon}
\email{george.privon@ufl.edu}

\shortauthors{Privon, Narayanan, and Dav\'e}

\title{On the Interpretation of Far-infrared Spectral Energy Distributions. I: The 850 $\mu$m Molecular Mass Estimator}
\shorttitle{The 850 {$\rm\mu m$} Molecular Mass Estimator}

\author[0000-0003-3474-1125]{G. C. Privon}
\affiliation{Department of Astronomy, University of Florida, 211 Bryant Space Sciences Center, Gainesville, FL 32611, USA}

\author[0000-0002-7064-4309]{D. Narayanan}
\affiliation{Department of Astronomy, University of Florida, 211 Bryant Space Sciences Center, Gainesville, FL 32611, USA}
\affiliation{University of Florida Informatics Institute, 432 Newell Drive,  Gainesville, FL 32611, USA}
\affiliation{Cosmic Dawn Center (DAWN), Niels Bohr Institute, University of Copenhagen, Juliane Maries vej 30, DK-2100 Copenhagen, Denmark}

\author[0000-0003-2842-9434]{R. Dav\'e}
\affiliation{Institute for Astronomy, Royal Observatory, Edinburgh EH9 3HJ, UK}
\affiliation{University of the Western Cape, Bellville, Cape Town 7535, South Africa}
\affiliation{South African Astronomical Observatory, Cape Town 7925, South Africa}

\begin{abstract}
We use a suite of cosmological zoom galaxy formation simulations and dust radiative transfer calculations to explore the use of the monochromatic $850~\mu m$ luminosity (\Le) as a molecular gas mass (\Mmol) estimator in galaxies between $0 < z < 9.5$ for a broad range of masses.
For our fiducial simulations, where we assume the dust mass is linearly related to the metal mass, we find that empirical \Le-\Mmol calibrations accurately recover the molecular gas mass of our model galaxies, and that the \Le-dependent calibration is preferred.
We argue the major driver of scatter in the \Le-\Mmol relation arises from variations in the molecular gas to dust mass ratio, rather than variations in the dust temperature, in agreement with the previous study of Liang et al.
Emulating a realistic measurement strategy with ALMA observing bands that are dependent on the source redshift, we find that estimating \Se from continuum emission at a different frequency contributes $10-20\%$ scatter to the \Le-\Mmol relation.
This additional scatter arises from a combination of mismatches in assumed \Tdust and $\beta$ values, as well as the fact that the SEDs are not single-temperature blackbodies.
However this observationally induced scatter is a sub-dominant source of uncertainty.
Finally we explore the impact of a dust prescription in which the dust-to-metals ratio varies with metallicity.
Though the resulting mean dust temperatures are $\sim50\%$ higher, the dust mass is significantly decreased for low-metallicity halos.
As a result, the observationally calibrated \Le-\Mmol relation holds for massive galaxies, independent of the dust model, but below \Le$\lesssim10^{28}$ erg s$^{-1}$ (metallicities $\log_{10}({\rm Z}/{\rm Z}_{\odot})\lesssim -0.8$) we expect galaxies may deviate from literature observational calibrations by $\gtrsim0.5$ dex.
\end{abstract}


\section{Molecular Reservoirs in Galaxies and Thermal Dust Emission}

Molecular gas has long been recognized as a key ingredient in galaxy evolution, largely through its consumption in star formation.
Accordingly, determining the mass of molecular gas reservoirs, \Mmol, has been pursued as an important property of galaxies.
The most common method of constraining the molecular gas content of galaxies is via observations of low-J transitions of CO, and converting to an equivalent \HH mass \citep[e.g.,][and references therein]{Bolatto2013,casey14a}.

Single-band continuum estimators of the molecular gas mass have been increasingly used to chart the evolution of the ISM content of galaxies at high redshift \citep[e.g.,][]{Scoville2014,Groves2015,Scoville2016,Schinnerer2016,Scoville2017,Bertemes2018}.
The continuum sensitivity of the Atacama Large Millimeter Array (ALMA) means dust measurements can be obtained rapidly for high-z sources \citep[e.g.,][]{Scoville2014}.
If these dust emission measurements can be reliably converted to molecular gas mass estimates, the sensitivity of ALMA would enable the study of the gas content of large samples of galaxies \citep[e.g,.][]{Scoville2016,Scoville2017}.
Generally, this takes the form of:
\begin{equation}
L_{\nu}= C T_{\rm dust} \kappa_{\nu} \frac{M_{\rm dust}}{M_{\rm gas}}M_{\rm gas} = \Upsilon M_{\rm gas}
\label{eq:DGR}
\end{equation}
Where C is an unknown proportionality constant, and $\Upsilon$ is empirically calibrated from existing observations.
We note that this approach differs from techniques which fit the full spectral energy distribution to derive \Mdust and then assume a dust-to-gas ratio to estimate \Mmol \citep[e.g.,][]{Magdis2012b}.

In a series of papers \citet{Scoville2014,Scoville2016} outlined a procedure for estimating molecular gas masses using long-wavelength continuum measurements, directly at $850~\mu m$ or by observing redshifted continuum emission from a higher-frequency rest-frame and down-converting it to $850~\mu m$.
They additionally calibrated this relationship empirically against massive galaxies at $z=0$ and $z\sim2$ which had both dust continuum ($850~\mu m$ or $500~\mu m$) and \CO{1}{0} measurements.
A similar calibration was also derived by using Planck observations of Milky Way molecular clouds.
\citet{Scoville2014} find a relationship of:
\begin{equation}
\frac{L_{\rm \nu,850}}{M_{\rm mol}} = \alpha_{\nu,850} = (6.7\pm1.7) \times 10^{19}~({\rm erg}~{\rm s}^{-1}~{\rm Hz}^{-1}~\rm{M}_{\odot}^{-1})
\label{eq:scoville}
\end{equation}

Note that $\alpha_{\rm \nu,850}$ differs from $\Upsilon$ in Equation~\ref{eq:DGR} because $\alpha_{\rm \nu,850}$ converts only to the molecular gas mass.
To perform this empirical calibration based on the galaxy sample \citet{Scoville2014,Scoville2016} assumed a CO to \HH conversion factor of $\alpha=6.5$ \Msun $($K km s$^{-1}$ pc$^{2})^{-1}$, including the contribution from He.
\citet{Hughes2017b} found a similar calibration using a sample of main sequence star forming galaxies across the redshift range $0.02 < z < 0.35$.
Additionally, \citet{Janowiecki2018} explored the observational systematics in the \Le-\Mmol relation for galaxies in the volume-limited Herschel Reference Survey.
The found that deviations from \Mmol-dust emission relations primarily correlate with the \HI/\HH fraction.

Here we test the proportionality between the thermal dust emission and the molecular gas mass by using a suite of hydrodynamic cosmological zoom simulations spanning a redshift range of $0\leq z \lesssim 9.5$.
These simulations are coupled with dust radiative transfer post-processing to explore the link between the dust emission and the galaxy molecular mass.
The advantage of this approach is that the \Mmol values derived from the synthetic \Le ``measurements'' (using the observationally derived $\alpha_{\rm \nu,850}$ conversion factor) can be directly compared with the molecular gas masses in the simulations.
Discrepancies can be further correlated with the known dust temperature (\Tdust), dust mass (\Mdust), and molecular gas to dust mass of the galaxies in the simulations.

We begin by describing the cosmological zoom simulations and radiative transfer post-processing (Section~\ref{sec:sims}), then apply the \Le mass estimation technique (Section~\ref{sec:L850}), before comparing to the intrinsic properties of the simulations (Section~\ref{sec:molcomp}), discussing the likely origins of scatter about the relationship (Section~\ref{sec:scatter}), and concluding with a brief discussion of the implications for high-redshift ALMA observations (Section~\ref{sec:ALMAobs}).
Throughout the paper we assume a Planck2013 cosmology \citep[$H_0=67.77$ \kms Mpc$^{-1}$, $\Omega_{\rm matter}=0.30713$;][]{PlanckXVI}.

\section{Numerical Methods}
\label{sec:sims}

Our aim is to use modeled submillimeter-wave flux densities
from galaxies at high-redshift to test how the dust continuum emission relates to the underlying
molecular gas mass and test observational \Mmol estimators.  To do this, we will model a sample of simulated
galaxies at high-redshift using the cosmological zoom technique, and
from those generate the synthetic broadband SEDs.  To do this, we will
follow \citet{olsen17a,narayanan18a} and \citet{Abruzzo2018},
and combine galaxies zoomed in on from the {\sc mufasa} cosmological
simulation series \citep{dave16a,dave17b,dave17a} with {\sc powderday}
dust radiative transfer \citep{narayanan15a,Narayanan2018b}.  In this section, we
will summarize these methods, though refer the readers to the
aforementioned works \citep[in particular][]{narayanan18a} for
further details.

\subsection{Cosmological Zoom Galaxy Formation Simulations}
We perform our galaxy formation simulations using the hydrodynamic
galaxy formation code {\sc gizmo}
\citep{hopkins15a,hopkins17a,hopkins17b}.  These simulations are
performed in meshless finite mass (MFM) mode, in which the cubic
spline kernel is used to define the volume partition between gas
elements (and, therefore, the faces over which the Riemann solver
solves the hydrodynamic equations).

The cosmological zoom technique isolates dark matter halos of interest
at a particular redshift, and re-simulates these at higher resolution.
Functionally, we first initialize the simulation
at $z=249$ using {\sc music} \citep{hahn11a}, where the initial
conditions are identical to those in the {\sc mufasa}
cosmological simulations \citep{dave16a}.  We then run a run a coarse
dark matter only simulation to $z=0$ with particle mass $7.8 \times
10^8 h^{-1} M_\odot$ in a $50 h^{-1}$ Mpc volume, and $512^3$ dark
matter particles.  From this dark matter simulation, we select model
halos to re-simulate at higher resolution, and with baryons included.

We identify these halos using {\sc caesar} \citep{thompson14a}, and
track all particles that fall within $2.5 \times r_{\rm max}$ of the
halo of interest back to $z=249$.  Here, $r_{\rm max}$ is the radius
of the farthest particle from halo center; this technique, while
computationally expensive, ensures that we have a $0\%$ contamination
rate of low-resolution particles in our final model halos.

For the purposes of this paper, we analyze $7$ model halos over a
broad range of final halo masses.  We list some relevant physical
properties of these halos in Table~\ref{table:simsum}.  The $4$ most massive
halos are selected from a $z=2$ snapshot, and only run to $z_{\rm
  final} \approx 2$, while the remaining $3$ were selected from a $z=0$
snapshot (and consequently run to $z_{\rm final} \approx 0$).  These
model halos range from approximately Milky Way mass at $z=0$ to
massive halos that may represent luminous dusty star forming galaxies
at high-redshift \citep{casey14a}.

\begin{deluxetable*}{lcccccc}
\tablecaption{Zoom Simulation Summary}
\tablehead{\colhead{Halo ID} & \colhead{$z_{\rm final}$} & \colhead{M$_{\rm DM}$($z=2$)} & \colhead{M$_{*}$($z=2$)} & \colhead{M$_{*}$($z_{\rm final}$)} \\
 & & \colhead{($\times10^{11}$ \Msun)} & \colhead{($\times10^{10}$ \Msun)} & \colhead{($\times10^{10}$ \Msun)} }\startdata
0 & 2.15 & 41\phantom{.0} & \nodata & 10.50 \\
5 & 2.05 & 63\phantom{.0} & \nodata & 10.32 \\
10 & 2.00 & 11\phantom{.0} & 8.82 & \phantom{1}8.82 \\
45 & 2.00 & 37\phantom{.0} & 1.02 & \phantom{1}1.02 \\
287 & 0.65 & \phantom{1}0.3 & 0.28 & \phantom{1}0.39 \\
352 & 0.00 & \phantom{1}0.9 & 0.47 & \phantom{1}3.42 \\
401 & 0.02 & \phantom{1}0.6 & 0.34 & \phantom{1}2.60 \\
\enddata
\tablecomments{Halo ID number, final redshift of zoom simulation, total halo mass at $z=2$, stellar mass within a 50 kpc box at $z=2$, and stellar mass within a 50 kpc box at the final redshift.}
\label{table:simsum}
\end{deluxetable*}

The baryonic zoom galaxy formation simulations are run with the {\sc
  mufasa} suite of physics \cite{dave16a}.  In short, stars form in
dense molecular gas according to a volumetric \citet{schmidt59a}
relation, with an imposed star formation efficiency per free fall time
of $\epsilon_* = 0.02$, as motivated by observations
\citep{kennicutt98b,kennicutt12a,narayanan08b,narayanan12a,hopkins13a}.
The molecular gas fraction is determined following
\citet{krumholz09a}, wherein the molecular gas fraction is tied to the
surface density of the gas, and its metallicity.
The gas surface density is computed using the Sobolev approximation, as described in \citet{dave16a} and uses kernel-smoothing of the nearest 64 neighbors.
The {\sc mufasa} cosmological simulations reproduce the observed $z=0$ \HH mass function and $f_{H_2}-M_*$ relation \citep{dave17a}.
When considering the \HH to CO conversion of \citet{narayanan12a}, the {\sc mufasa} simulations also reasonably reproduce the CO (1--0) luminosity function out to $z\sim2$.

Alternate \HH prescriptions exist, including a modified KMT model \citep{Krumholz2013a} and models dependent on the gas-to-dust ratio and interstellar radiation field \citep[e.g.,][]{Gnedin2011,Gnedin2014b}.
\citet{Lagos2015} explored the differences of the \citet{Krumholz2013a} and \citet{Gnedin2011} models on the implied \HH properties of the EAGLE simulation \citep{Schaye2015}.
\citet{Popping2014} also explored pressure \citep{Blitz2006} and metallicity-based \citep{Gnedin2011} \HH prescriptions in semi-analytic models.
Both studies found that the $z=0$ \HH mass function was reasonably well reproduced by all the models.
Discrepancies in \HH masses between models were most significant in low-metallicity galaxies \citep[Z$<0.5$Z$_{\odot}$;][]{Lagos2015} or low mass halos \citep[M$_{\rm halo}<10^{10}$ M$_{\odot}$;][]{Popping2014}.

In principle, variations in \HH could be explored for these zooms by re-evaluating them in post-processing and choosing an alternate sub-grid model.
However, this would introduce inconsistencies in the analysis, compromising a fair comparison.
Changes in the \HH model would propagate to changes in the star formation histories, which in turn would affect the metallicity.
The modified metallicity and star formation history would further result in changes in the dust masses and the dust temperatures.
These changes are not straightforward to estimate and would obviate a clear interpretation of the impact of varying \HH prescriptions, so we have not attempted to do so here.
This is a problem which is likely best addressed with a future study running new suites of zoom simulations with different \HH prescriptions to perform an internally consistent study.

We track the evolution of $11$ elements: H, He, C, N, O, Ne, Mg, Si,
S, Ca and Fe.  We draw SNe type 1a yields from \citet{iwamoto99a},
assuming $1.4 M_\odot$ of mass returned into the ISM per supernova.
Following \citet{dave16a}, type SNe type II yields are derived from
the \citet{nomoto06a} prescription, though we reduce these by $50\%$
in order to match mass-metallicity constraints at high-redshift
\citep{dave11a}.  The dust yields from AGB stars derive from the
\citet{oppenheimer08a} lookup tables.

Feedback from massive stars are included as a decoupled two phase wind
scheme in the {\sc mufasa} wind model.  Here, the stellar winds have a
probability for ejection that is a fraction of the star formation rate
probability.  This fraction derives from scaling relations from the
Feedback in Realistic Environments project
\citep{muratov15a,hopkins14b,hopkins17b}.  Here, the ejection velocity
depends on the galaxy circular velocity, which is determined on the
fly using a fast friends of friends finder.  AGB and Type Ia
supernovae winds are included, following \citet{bruzual03a} tracks
with a \citet{chabrier03a} initial mass function. These simulations
broadly agree with observational constraints of the SFR-$M_*$
relation, and $M_*-M_{\rm halo}$ relation \citep{Abruzzo2018}

\subsection{Dust Radiative Transfer}

With our model galaxies in hand, we now turn to generating
their synthetic broadband SEDs.  We do this using the publicly
available {\sc powderday} simulation
package \citep{narayanan15a,Narayanan2018b}, that wraps {\sc fsps} for the stellar
population synthesis \citep{conroy09b,conroy10b,conroy10a} with {\sc
yt} for grid generation \citep{turk11a}, and {\sc hyperion} for dust
radiative transfer \citep{robitaille11a,robitaille12a}.

Functionally, we cut a $50$ kpc box around the model central galaxy in
each snapshot, and build an adaptive grid with an octree memory
structure.  We begin with one cell encompassing all gas particles in
this box, and subdivide each cell into octs until a threshold number
of particles $n_{\rm thresh} = 64$ is reached in each cell.

The SEDs for the star particles within each cell are generated with
{\sc fsps}\footnote{Functionally, we use the python bindings for {\sc
fsps}, {\sc python-fsps}; \url{http://dfm.io/python-fsps/current/}}
using the stellar ages and metallicities as returned from the
cosmological simulations.  For these, we assume a \citet{kroupa02a}
stellar IMF and the Padova stellar
isochrones \citep{marigo07a,marigo08a}.  These stellar SEDs provide
the input spectrum that transfers through the dusty ISM of the galaxy.

For our fiducial simulations, the dust mass of each cell is assumed to be tied to the metal mass by
a constant fraction: $M_{\rm dust} = 0.4 \times M_{\rm metal}$.  This
is motivated by observations of both local galaxies and those at
high-redshift \citep{dwek98a,vladilo98a,watson11a,Sandstrom2013}.
We also briefly explore a prescription in which the dust-to-metals ratio is a smoothly varying function of the metallicity (Section~\ref{sec:dustformation}, Appendix~\ref{app:RR}), motivated by \citet{Remy-Ruyer2014}.
In both cases, the dust is
modeled as the carbonaceous-silicate \citet{draine07a} model that
follows the \citet{weingartner01a} size distribution, and
the \citet{draine03a} renormalization relative to Hydrogen.  We assume
$R_{\rm V} \equiv A_{\rm V}/E\left(B-V\right) = 3.15$.  This dust is
assumed to uniformly fill each cell.  We have performed resolution
studies to ensure that reducing $n_{\rm thresh}$ does not change our
results appreciably.
Unless otherwise noted, our discussions refer to the fiducial dust model.
In this paper we use ``metallicity'' to refer to the total metals plus dust.

The radiation from the stars is propagated throughout the grid in $3$
dimensions using {\sc hyperion}.  The radiative transfer occurs in a
Monte Carlo fashion, and employs the \citet{lucy99a} equilibrium
algorithm to determine the equilibrium condition between the radiation
field and dust temperature.  In this process, we emit radiation from
all of the stellar sources, and this radiation is absorbed, scattered,
and remitted from each cell.  We iterate on this process until the
energy absorbed by $99\%$ of the cells has changed by less than $1\%$.

It is important to note that we do not include any sub-resolution
models for birth
clouds \citep[e.g.][]{groves04a,jonsson10a,narayanan10a}.  The
compactness and covering fraction of these birth clouds tends to be
free parameters, and can have a significant impact on the final dust SED,
depending on the final parameter choices.
We discuss the impact of a birth cloud model in Appendix~\ref{app:birthcloud}.

New to these calculations \citep[as compared to][]{narayanan18a}, we
include the effects of the cosmic microwave background (CMB).  As we
will show, in some scenarios this can make an impact on our results \citep[see also][]{daCunha2013b}.
We include the CMB as an additional energy density term in each cell
as the energy absorbed per unit dust mass in each cell (i.e. $\epsilon
= \int \kappa_\nu B_\nu d\nu$ erg/s/g, where $\kappa_\nu$ is the dust
absorption opacity, and $B_\nu$ is the Planck function).  The CMB
temperature is simply $T_{\rm CMB} = 2.73(1+z)$ K, where $z$
is the redshift of the snapshot.

We note that the values for the radiative transfer parameters
described thus far are constrained by observed data and we did
not engage in any tuning of the treatment of dust properties in the
simulations.  The net result of our radiative transfer calculations is
a broadband SED, from $\lambda = 912$ \AA--3 mm (3287 THz -- 100 GHz).
It is these SEDs that we analyze for the remainder of this paper.

\begin{figure}
\includegraphics[width=0.45\textwidth]{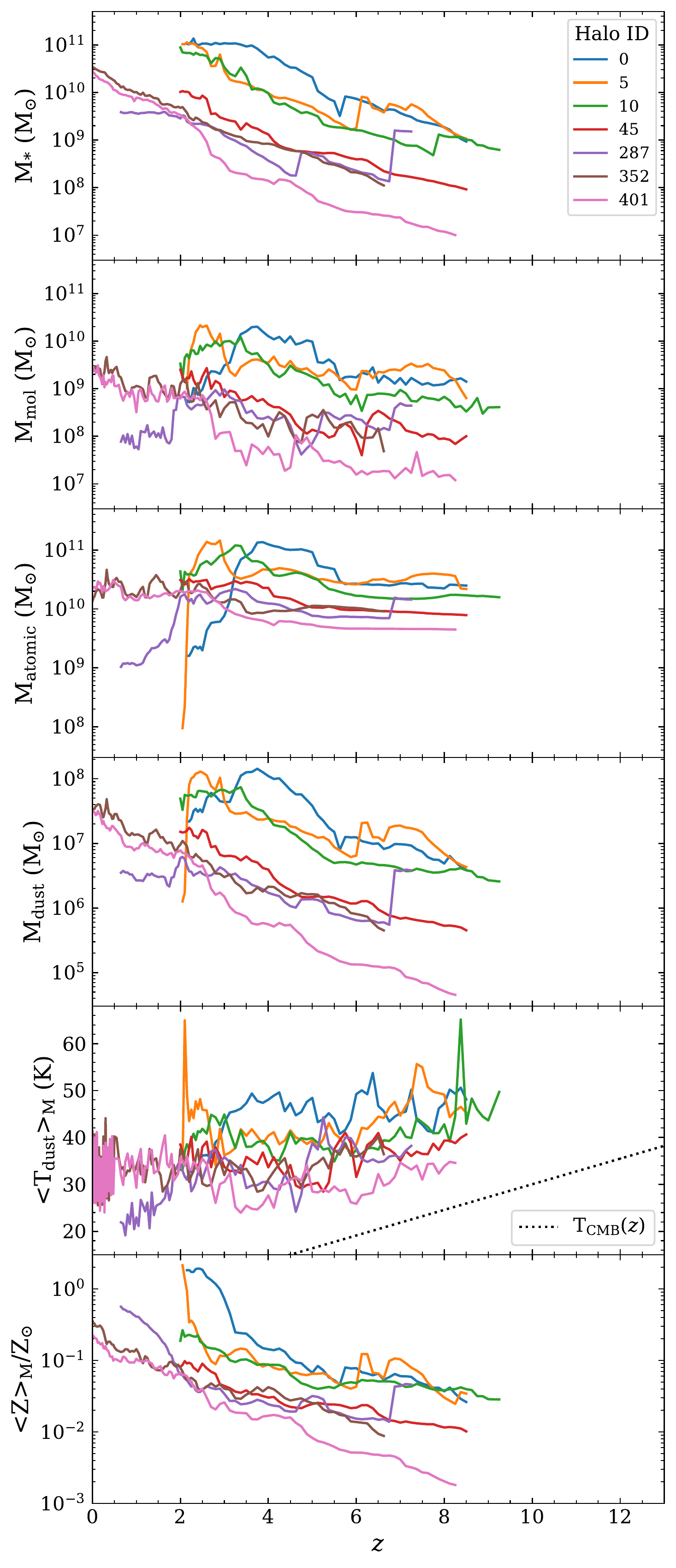}
\caption{The evolution of various physical properties of the simulated galaxies, as directly measured from the snapshots and subsequent post-processing.
From top to bottom: stellar mass (M$_{*}$), molecular gas mass (\Mmol) including He, atomic gas mass (M$_{\rm atomic}$), dust mass (\Mdust), molecular mass-weighted mean dust temperature($\langle T_{\rm dust}\rangle_M$), and mean gas phase metallicity weighted by the total gas mass ($\langle Z\rangle_M$).
In the panel showing \Tdust, the black dotted line shows the CMB temperature as a reference.
All quantities in this and subsequent plots are computed within a 50 kpc by 50 kpc box centered on the center of mass of the most massive galaxy in the high-resolution dark matter halo.}
\label{fig:simproperties}
\end{figure}

In Figure~\ref{fig:simproperties} we show the evolution of the physical properties of each simulated halo, including the stellar mass (M$_{*}$), molecular gas mass (\Mmol) including He, atomic gas mass (M$_{\rm atomic}$), dust mass (\Mdust), molecular mass-weighted mean dust temperature($\langle T_{\rm dust}\rangle_M$), and mean gas phase metallicity weighted by the total gas mass ($\langle Z\rangle_M$).
All quantities are computed inside a box 50 kpc across, centered on the center of mass of the central galaxy in the halo.
The dust mass appears to broadly track \Mmol and the metallicity, and is roughly independent of the atomic gas mass (here taken to include everything which is not in molecular form).
The dust temperatures begin high ($\sim30-50$ K), reflecting a combination of the intrinsic heating from star formation and low dust masses along with additional heating from the CMB (Appendix~\ref{app:CMB}), eventually decreasing to typical low-redshift values of $\sim30$ K.
In the remainder of the paper we explore the thermal continuum emission from the dust and compare it with \Mmol, $\langle T_{\rm dust}\rangle_M$, $\langle Z\rangle_M$, and \Mdust/\Mmol, to assess the precision of \Le as an estimator for \Mmol.

\section{Single-band Mass Indicators}
\label{sec:L850}

Here we explore the application of the \Le mass estimator to the SEDs
generated from the cosmological zoom simulations, using the same
assumptions made for interpretation of observations.  We seek to
answer two questions:
\begin{enumerate}
    \item{Do we recover a close link between a galaxy's intrinsic \Le and its molecular gas mass \Mmol using these realistic simulations?}
    \item{If there is a good \Le-\Mmol relation, what impact do realistic observing techniques, namely band-correction of fluxes, have on our ability to accurately recover \Mmol?}
\end{enumerate}

To answer the first question we explore an ideal scenario where the rest-frame $850~\mu m$ emission can be measured to directly compute \Led, then converting this to a molecular mass using the \citet{Scoville2016} calibration (Section~\ref{sec:ideal}).
This will directly probe the intrinsic link between the 850~$\mu$m continuum and \Mmol.

To answer the second question we investigate a more observationally realistic case where observations performed are of emission with $\nu_{\rm rest} > \nu_{850}$.
These observations must then be converted to an equivalent $850~\mu m$ flux by assuming optically thin emission, a dust emissivity index $\beta$, and a (mass-weighted) dust temperature, $\langle T_{\rm dust}\rangle_{\rm M}$ (Section~\ref{sec:redshifted}).
This latter approach results in an inferred luminosity, \Lei.
Comparison of these two cases will enable us to separate any offsets introduced by the band conversion (i.e., $\nu_{\rm rest} > \nu_{850}$) from scatter in the direct \Le--\Mmol relation.
In Table~\ref{table:symbols} we briefly summarize commonly used symbols in this paper.

Note that we do not consider the effect of measurement errors on the fluxes or the observational contrast against the CMB \citep{daCunha2013b}, though heating of the dust by the CMB is included.
Our aim is to test the physical validity of the link between \Le and \Mmol under ideal observational conditions; measurement noise and/or contrast issues will introduce additional bias or scatter beyond what we find here.
Exploration of the influence of non-detections and noise-induced scatter will be considered in Paper II, when we investigate the link between multi-frequency dust SED measurements and the properties of dust in the simulations.

\begin{deluxetable*}{lll}
\tablecaption{Frequently Used Symbols}
\tablehead{\colhead{Symbol} & \colhead{Units} & \colhead{Definition}}
\startdata
\Le & erg s$^{-1}$ Hz$^{-1}$    & 850~$\mu$m monochromatic luminosity \\
\Led & erg s$^{-1}$ Hz$^{-1}$    & 850~$\mu$m monochromatic luminosity, measured directly from the $850~\mu$m flux density \\
\Lei & erg s$^{-1}$ Hz$^{-1}$    & 850~$\mu$m monochromatic luminosity, inferred from a higher $\nu_{rest}$ observation \\
$\Gamma_{\rm RJ}$ & \nodata & Correction factor for flux extrapolation along RJ tail, see \citet{Scoville2016}\\
\Mmol & M$_{\odot}$ & Molecular gas mass (including He)\\
M$_{\rm mol,850}$ & M$_{\odot}$ & Molecular gas mass (including He) derived from \Le \\
M$_{\rm mol, true}$ & M$_{\odot}$ & Molecular gas mass (including He) in the hydrodynamic simulations\\
$\alpha_{\nu,850}$ & erg s$^{-1}$ Hz$^{-1}$ M$_{\odot}^{-1}$ & Constant conversion factor between \Le and \Mmol, from \citet{Scoville2016}\\
$\alpha_{\nu,850}($\Le$)$ & erg s$^{-1}$ Hz$^{-1}$ M$_{\odot}^{-1}$ & \Le-dependent conversion factor between \Le and \Mmol, from \citet{Scoville2016}\\
\enddata
\tablecomments{A summary of some symbols used in the text along with their units and definitions.}
\label{table:symbols}
\end{deluxetable*}

\subsection{Ideal Case: Direct $850~\mu m$ measurement}
\label{sec:ideal}

\begin{figure*}
\includegraphics[width=0.5\textwidth]{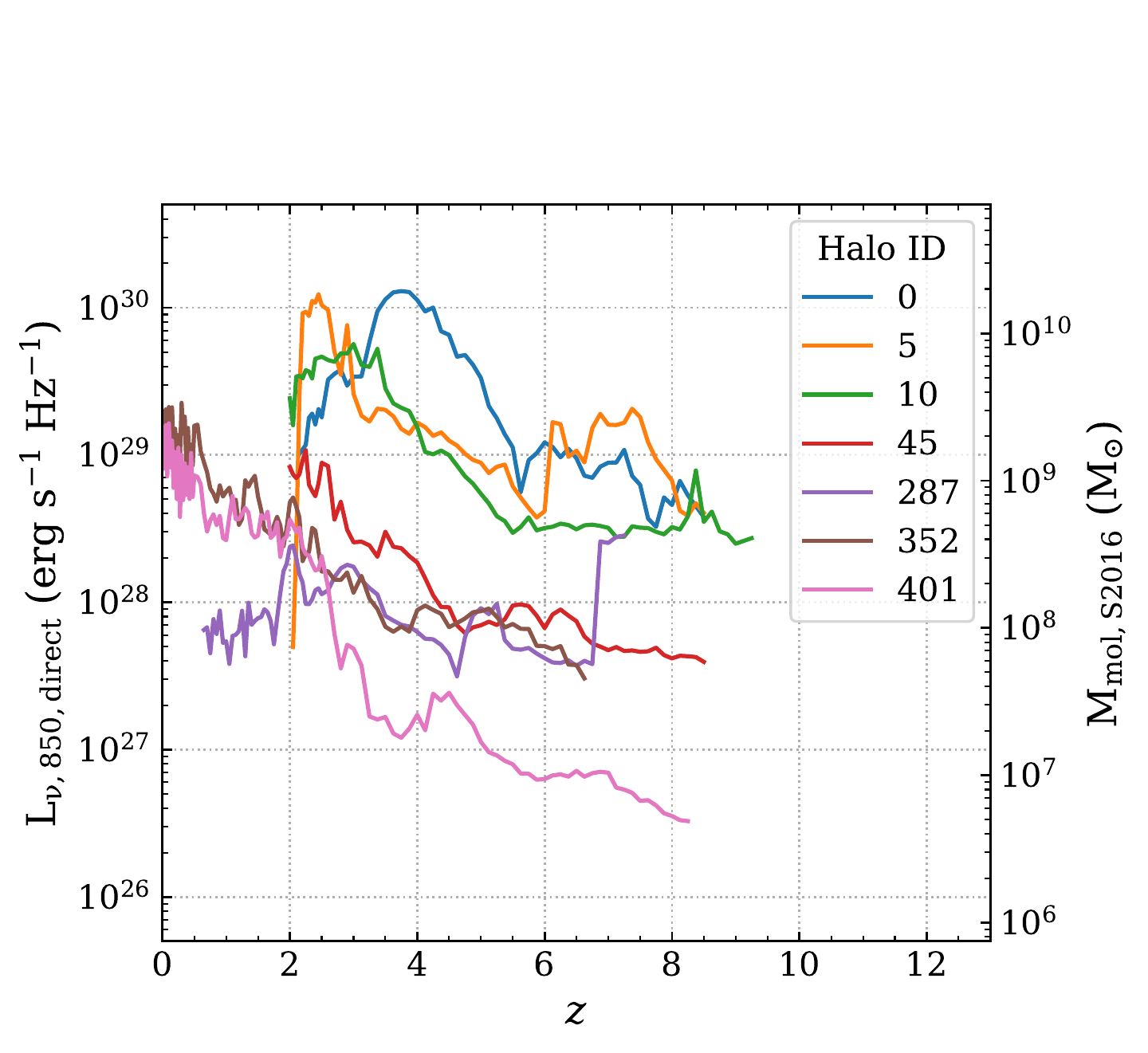}
\includegraphics[width=0.5\textwidth]{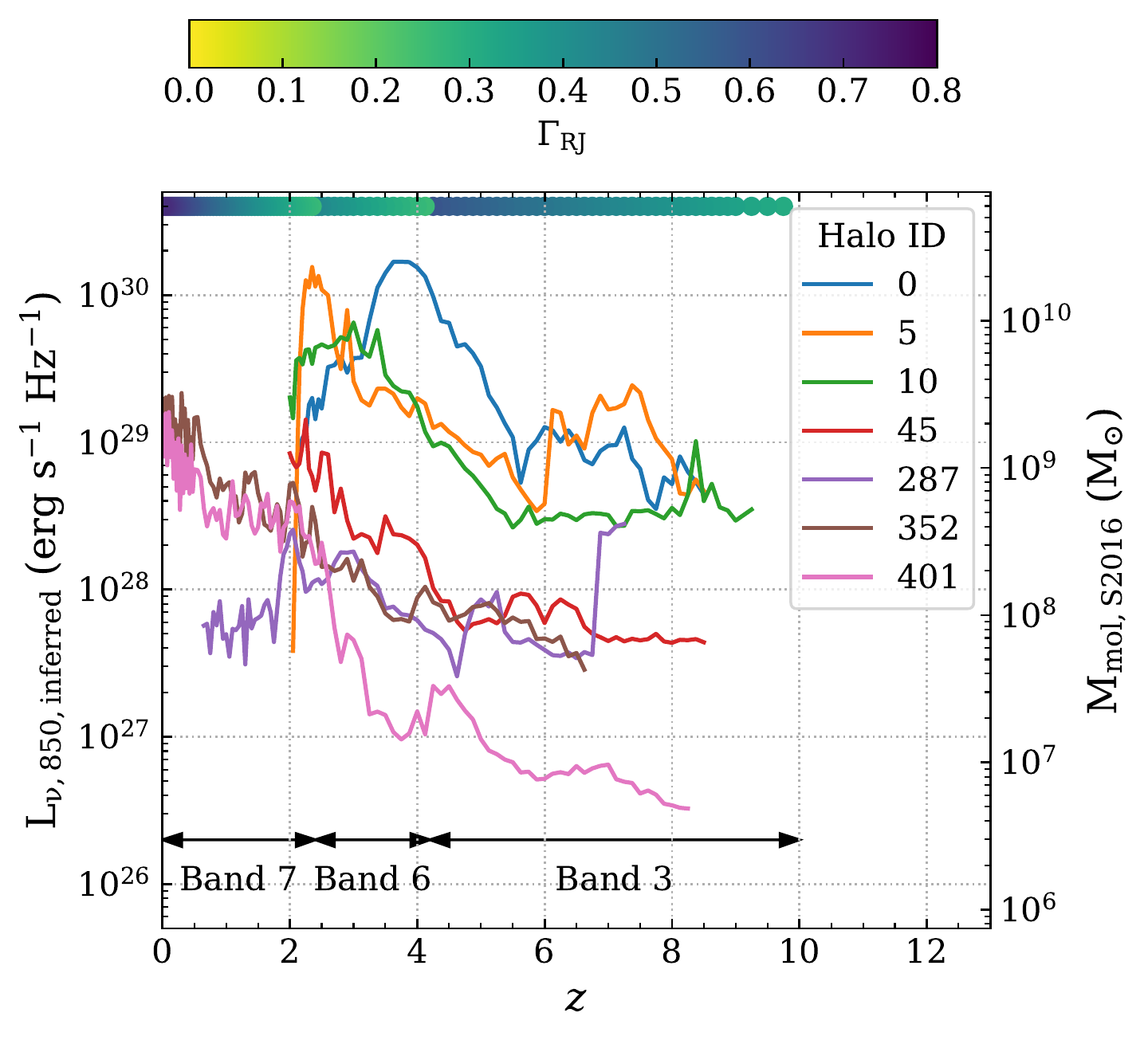}\\
\includegraphics[width=0.5\textwidth]{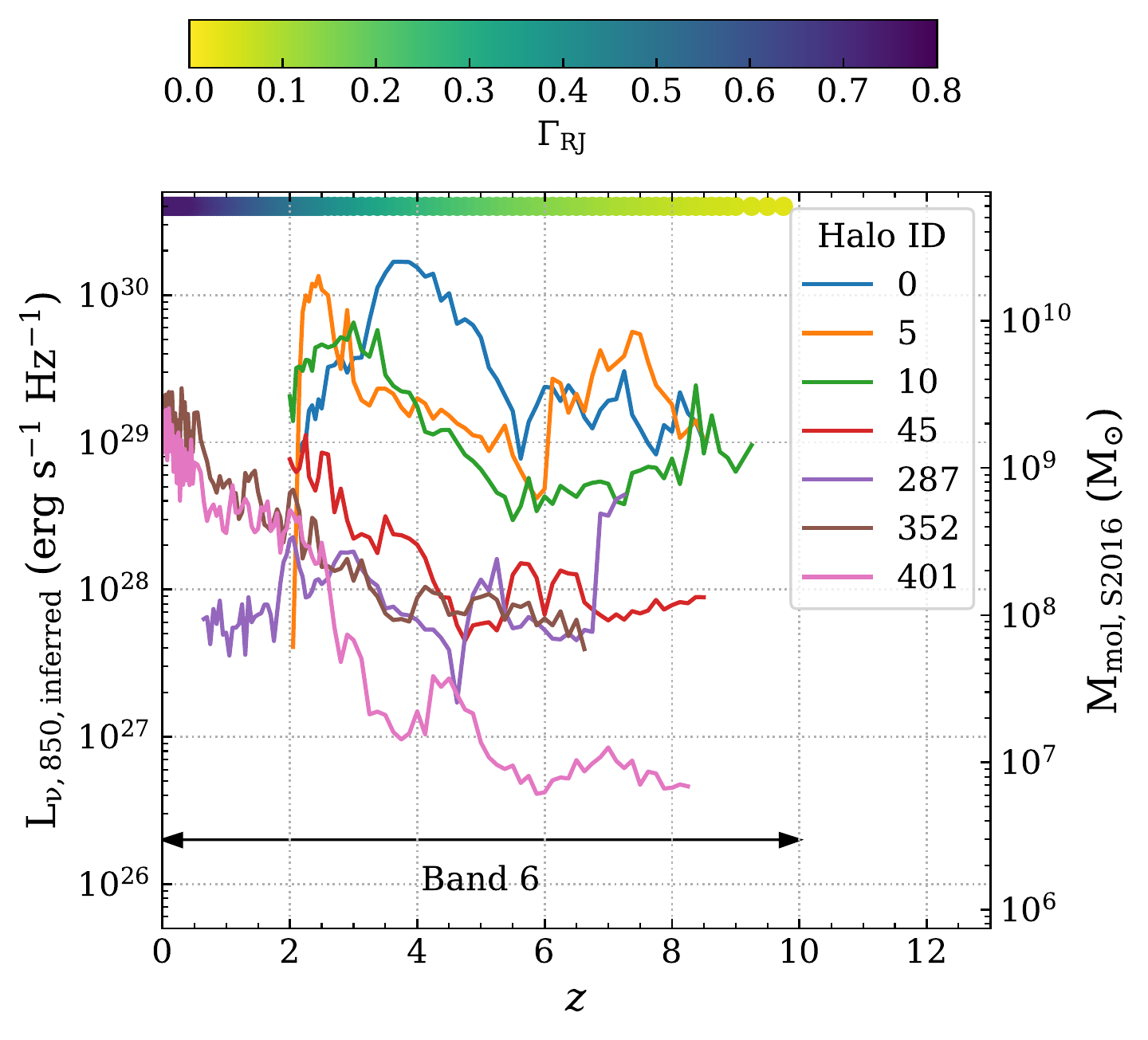}
\includegraphics[width=0.5\textwidth]{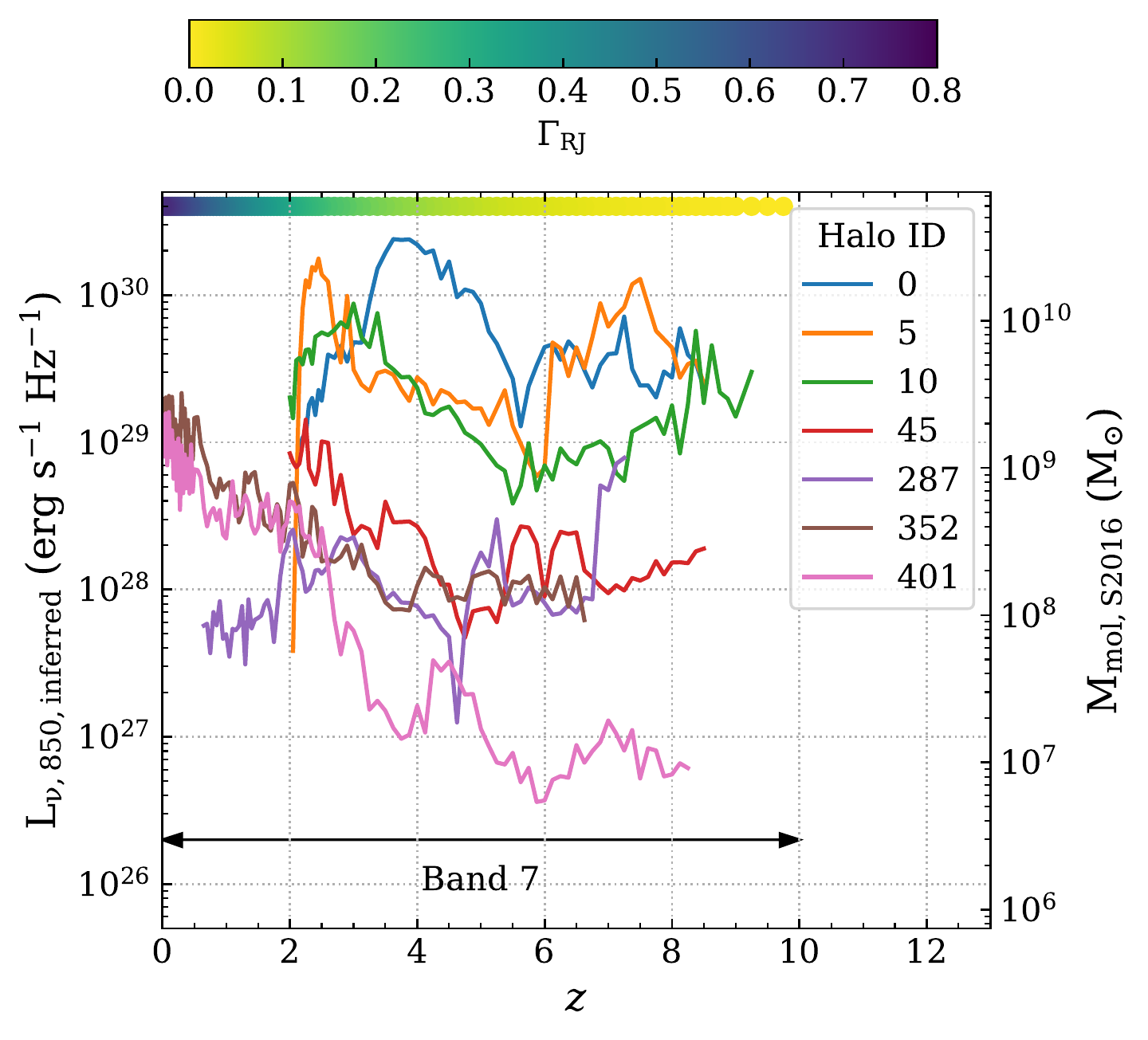}
\caption{The monochromatic $850~\mu m$ luminosity, \Le as a function of redshift for the individual galaxy halos of our zoom simulations.
The upper left panel shows \Led, for the ideal case where the rest-frame $850~\mu m$ emission can be directly observed.
The upper right panel shows \Lei, where \Le is inferred from a higher rest-frequency observation.
The lower row shows \Lei when only Band 6 (left) or only Band 7 observations (right) are available.
In all panels aside from the upper left, the arrows mark the ranges for the ALMA band used to determine the observed frequency.
The colored stripe near the top of the figure shows the $\Gamma_{\rm RJ}$ correction factor applied at a given redshift \citep[see Appendix A of][]{Scoville2016}, assuming \Tdust$=25$ K and $\beta=1.8$.
Note that the upper left panel does not require any assumptions regarding \Tdust or $\beta$.
The right ordinate of all panels shows the inferred molecular gas mass using the \citet{Scoville2016} calibration with a constant $\alpha_{850}$.
At high redshifts, Band 6 and 7 observations result in overestimates of of \Le, of 0.5 dex (see Section~\ref{sec:bandconv}).}
\label{fig:L850-z}
\end{figure*}

For each snapshot from each cosmological zoom simulation we directly extract \Se from the resulting SED, functionally equivalent to an $850~\mu m$ rest-frame observation being possible for all redshifts.
We then directly compute \Led using the (known) luminosity distance at the redshift of the snapshot.
The computed \Led for each halo as a function of redshift is shown in Figure~\ref{fig:L850-z} (left), as well as the \Mmol implied by the \citet{Scoville2016} calibration (Equation~\ref{eq:scoville}).
The halos collectively span an implied range in molecular gas mass from $10^6$ to $\sim2\times10^{10}$ \Msun.

\subsection{Realistic Case: Conversion of Redshifted Observations to $850~\mu m$}
\label{sec:redshifted}

Directly observing the rest-frame $850~\mu m$ observation is often not feasible, so we further emulate the common observational tactic of observing higher-frequency emission and using this to estimate \Se (and hence \Led) by assuming a dust emissivity spectral ($\beta$) index and a dust temperature (\Tdust).
The dust temperature enters through the Rayleigh-Jeans correction factor, $\Gamma_{\rm RJ}$ \citep{Scoville2016}.
For consistency with quantities typically applied to observations we assume $\beta=1.8$ and \Tdust$=25$ K.

\subsubsection{Redshift-dependent Band Selection}

It is desirable to select an observing band that balances the competing requirements that this correction factor does not become too large (i.e., $\nu_{\rm rest}$ is still on the RJ tail) and where the flux density is as high as possible (to optimize integration times and detectability).
Though we do not consider detection statistics and measurement noise here, we select continuum bands which would be reasonable observational choices \citep[see e.g.,][]{Scoville2014}.
Specifically we use frequencies corresponding to ALMA observations at Band 7 (353 GHz) for $z\leq2.4$, Band 6 (233 GHz) for $2.4<z<4.2$, and Band 3 (97.5 GHz) for $z\geq4.2$ (Table~\ref{table:ALMA}).

\begin{deluxetable}{lcc}
\tablecaption{Fiducial ALMA Observing Parameters}
\tablehead{\colhead{Band} & \colhead{$\nu_{\rm obs}$} & \colhead{RMS in 1 hour}\\
 & \colhead{(GHz)} & \colhead{(\mJybeam)}}
\startdata
3 & \phantom{0}97.5 & 0.011 \\
6 & 233.0 & 0.013 \\
7 & 343.5 & 0.030
\enddata
\tablecomments{These are the ALMA bands and observing frequencies assumed for deriving \Lei (Section~\ref{sec:redshifted}), obtained from the ALMA Cycle 6 standard continuum frequencies.
The sensitivities were obtained with the ALMA Cycle 6 sensitivity calculator and assume one hour on-source with 7.5 GHz bandwidth.
Sensitivity numbers are only employed in estimating the detection prospects for the high-redshift halos (Section~\ref{sec:ALMAobs}).
}
\label{table:ALMA}
\end{deluxetable}

Similar to the ideal case (Section~\ref{sec:ideal}), we extract $S_{\rm \nu,obs}$ at the frequency prescribed above for the redshift of each individual snapshot.
We then convert this to an estimated \Se following the procedure outlined in \citet{Scoville2016}.
In Figure~\ref{fig:L850-z} (right) we show the \Lei values inferred using this technique, for each halo as a function of $z$.
Unsurprisingly, the general behavior of the inferred \Led values closely tracks that of the directly measured \Led (Figure~\ref{fig:L850-z}, left) with the same implied range in \Mmol.
Encouragingly, the differences appear by eye to be relatively minor and we will assess the accuracy of the band conversion in Section~\ref{sec:bandconv}.

\subsubsection{Redshift-independent Band Selection}

Practical considerations (i.e., availability of data) may result in observations which are not ``optimal'' in terms of minimizing the impact of the $\Gamma_{RJ}$ correction factor.
In order to assess the recoverability of \Le under these conditions we also explore scenarios in which either ALMA Band 6 or ALMA Band 7 (Table~\ref{table:ALMA}) are the only observations available at all redshift.
In these cases the disagreement between \Lei and \Led can be much more significant (Section~\ref{sec:bandconv}).

\section{Recovery of \Mmol using \Le}
\label{sec:molcomp}

With these two ``measurements'' of \Le in hand, we now explore the correspondence of the implied \Mmol values with the molecular gas masses of the simulations.
In Figure~\ref{fig:S2016comp} we plot the \Mmol values from our simulations against \Le.
We also show the observed quantities used by \citet{Scoville2016} to calibrate their relationship.
The \Mmol values from \citet{Scoville2016} were computed from \Lprime{CO}{1}{0} observations and $\alpha_{\rm CO}=6.5$ \Msun $($K km s$^{-1}$ pc$^{2})^{-1}$.
For our simulations we use the \Mmol values in the snapshots, which, as a reminder, are computed in the cosmological galaxy formation simulations utilizing the \citet{krumholz09a} model for the HI-H$_2$ phase balance in clouds (see Section~\ref{sec:sims}).

\subsection{Empirical Calibrations of $\alpha_{850}$}

\begin{figure*}
\centering
\includegraphics[width=0.45\textwidth]{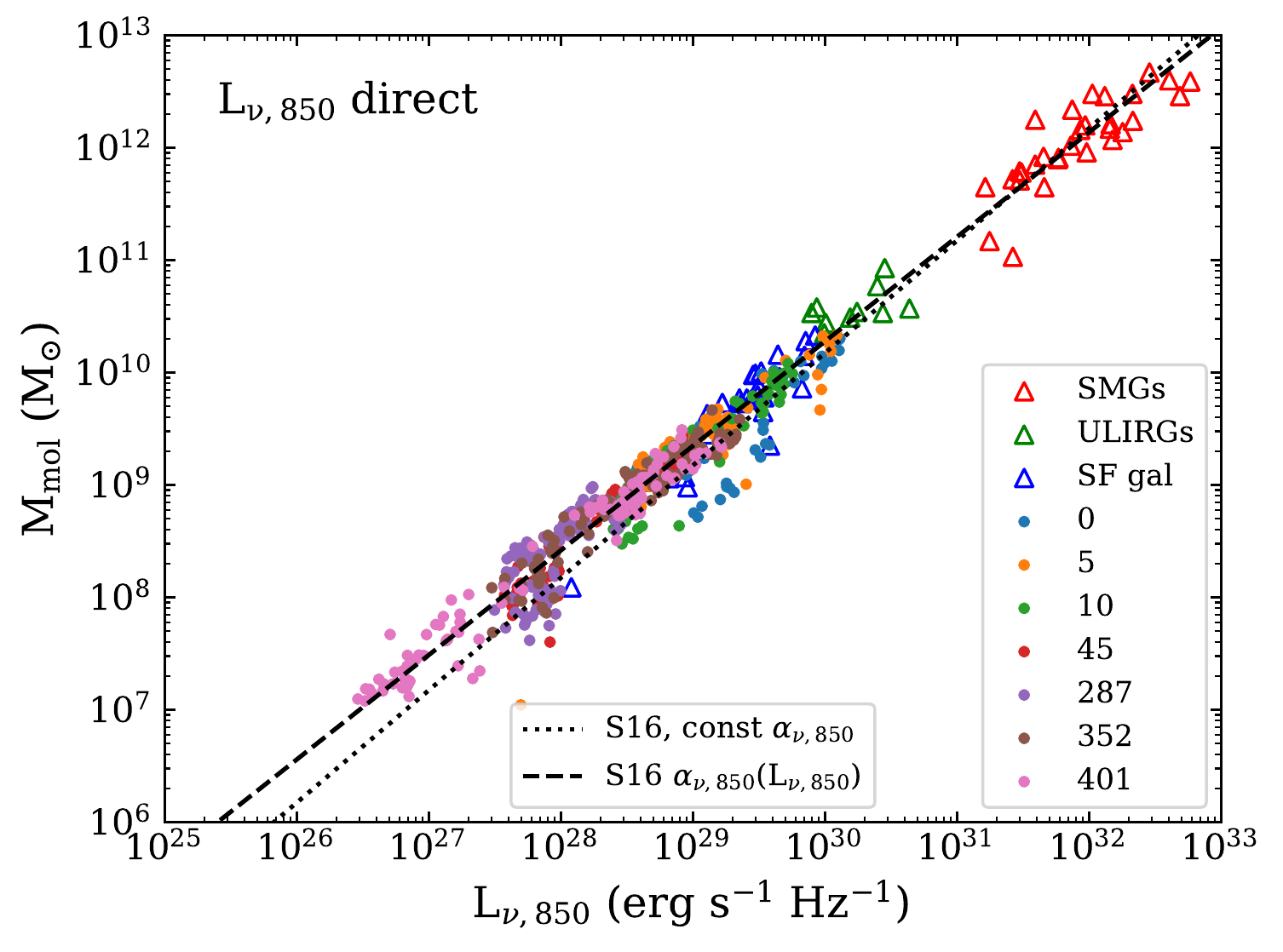}
\includegraphics[width=0.45\textwidth]{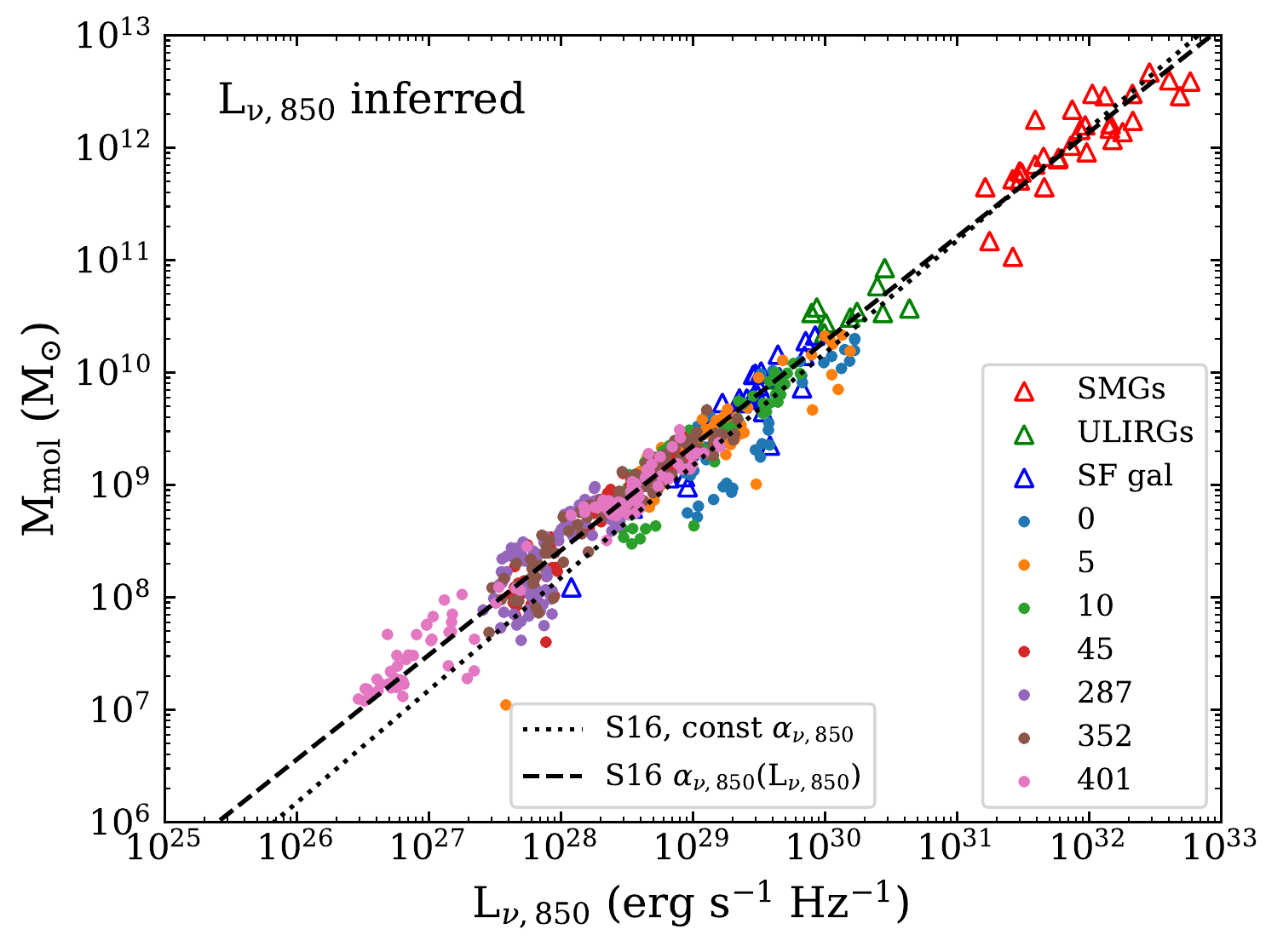}
\caption{A comparison of the observed galaxy data and our simulations in the \Le-\Mmol plane.
The sample used by \citet{Scoville2016}, to arrive at their \Le-\Mmol conversion is marked with triangles.
The dots mark the locations of our simulation snapshots, where \Le is obtained as described in Section~\ref{sec:L850} and \Mmol is directly measured from the simulation snapshots.
The dashed line corresponds to the median \Le/\Mmol relation of \citet{Scoville2016} while the dotted line shows their best fit.
The left panel shows \Led measured directly (Section~\ref{sec:ideal}) and the right panel shows \Lei determined from higher rest-frequency observation and down-converted (Section~\ref{sec:redshifted}).
The molecular masses for the \citet{Scoville2016} data were obtained by them from \Lprime{CO}{1}{0} by assuming $\alpha_{\rm CO}=6.5$ \Msun $($K km s$^{-1}$ pc$^{2})^{-1}$.
Note that many of the submillimeter galaxies (SMGs; red triangles) may be lensed; following \citet{Scoville2016} we are plotting their apparent luminosities here and not correcting for lensing.
The luminosity-dependent calibration more closely tracks the results of our simulations.
}
\label{fig:S2016comp}
\end{figure*}

\citet{Scoville2016} provide two fits to their observed data: a constant $\alpha_{\nu,850}$ ($=6.7\times10^{19}$ erg s$^{-1}$ Hz$^{-1}$ \Msun$^{-1}$) and a conversion with weak \Le dependence:

\begin{equation}
\alpha_{\nu,850}({\rm L}_{\nu})=6.2\times10^{19}\left ( \frac{{\rm L}_{\nu}}{10^{31} {\rm~erg~s}^{-1}~{\rm Hz}^{-1}}\right )^{0.07}~{\rm erg~s}^{-1}~{\rm Hz}^{-1}~{\rm M}_{\odot}^{-1}
\end{equation}
In Figure~\ref{fig:S2016comp} we overplot the constant $\alpha_{\nu,850}$ as well as the luminosity-dependent $\alpha_{\nu,850}($L$_{\nu})$.

\begin{figure*}
\centering
\includegraphics[width=\textwidth]{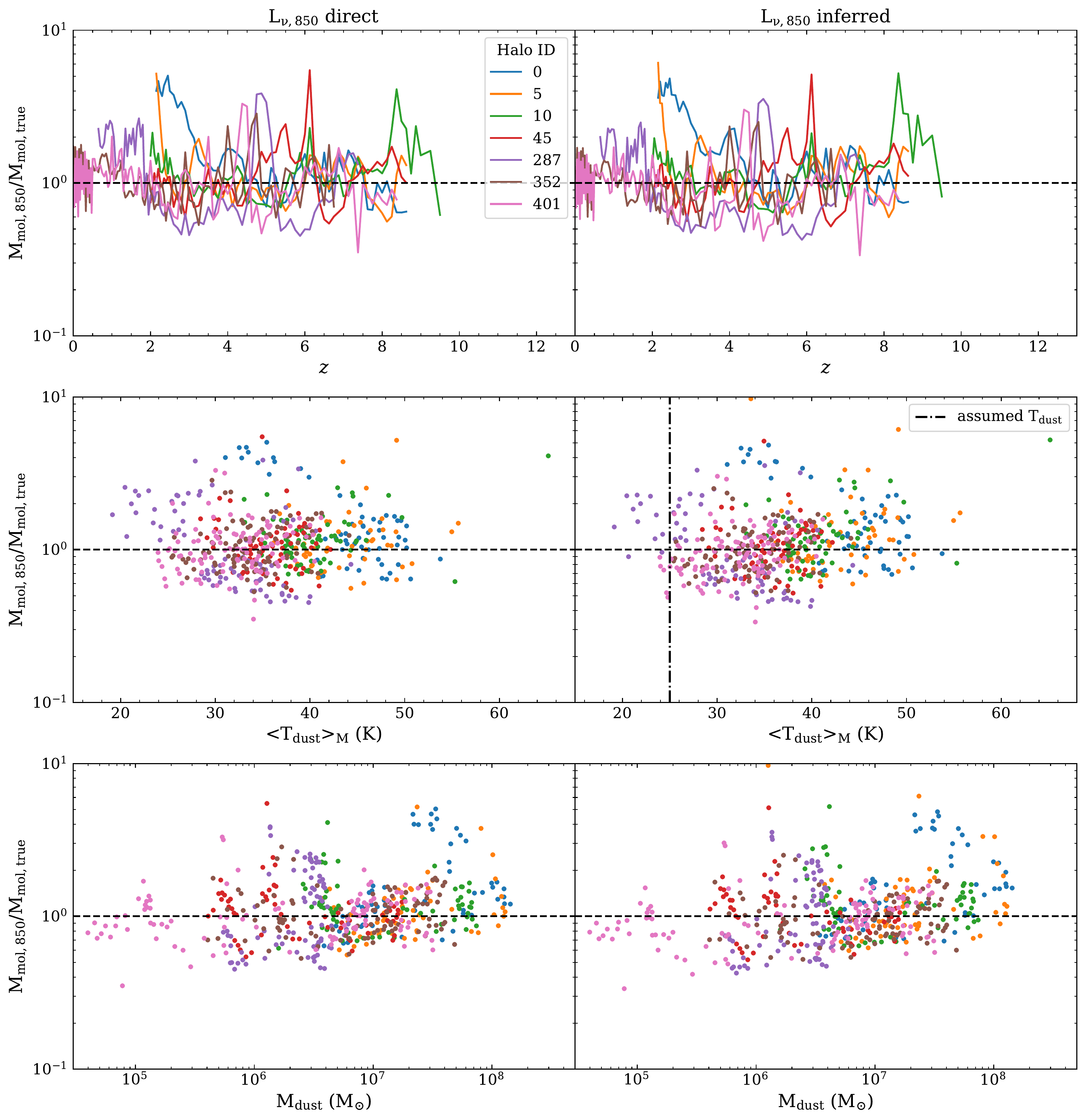}
\caption{A comparison of \Mmol values directly measured from the simulations with \Mmol derived using \Le and the \citet{Scoville2016} luminosity-dependent calibration.
The left column shows \Mmol estimates using \Led values (Section~\ref{sec:ideal}) and the right column shows \Mmol estimates using \Lei inferred from a redshifted observation (Section~\ref{sec:redshifted}).
In both cases, they are compared to the \Mmol of the simulation snapshots (i.e., \emph{not} derived from the \Le relation).
From top to bottom we compare with: redshift, the mass-weighted mean \Tdust, and \Mdust
In all panels, the horizontal black dashed line shows equality between \Mmole and \Mmolt.
The scatter about the line of equality correlates most strongly with the galaxy-integrated gas to dust ratio.
We emphasize that in obtaining \Mmol from the \Led values (left column) it is not necessary to make any assumptions for \Tdust or $\beta$.
}
\label{fig:mmolcomp}
\end{figure*}

\begin{figure*}
\addtocounter{figure}{-1}
\centering
\includegraphics[width=\textwidth]{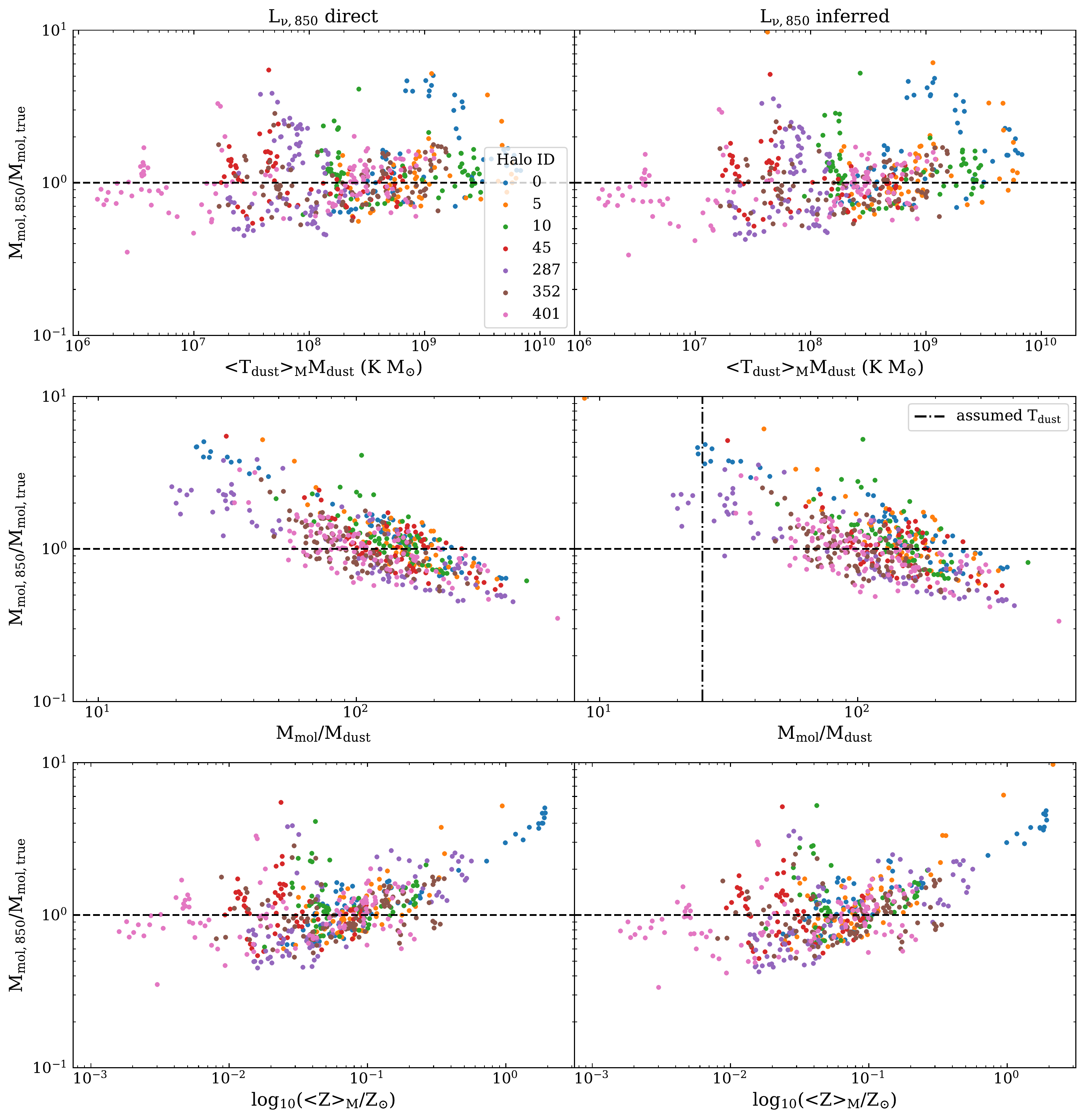}
\caption{Continued, showing from top to bottom:
the product $\langle {\rm T}_{\rm dust}\rangle_{\rm M}{\rm M}_{\rm dust}$, and the molecular gas to dust mass ratio.}
\end{figure*}

In both panels of Figure~\ref{fig:S2016comp} we find a good correlation between \Le and \Mmol for our simulated galaxies.
Independent of how we obtain \Le from the simulations, our snapshots more closely track the \Le-dependent $\alpha_{\nu,850}$ conversion.
The difference between the calibrations only exceeds a factor of two when ${\rm L}_{\nu,850}<1.5\times10^{27}$ erg s$^{-1}$ Hz$^{-1}$.
However this underestimate is systematic and should be considered when exploring samples with a range of intrinsic \Le values.
We argue that the observationally calibrated \Le-\Mmol relation successfully recovers molecular masses for our fiducial case of a constant dust-to-metals ratio.

\section{Origin of the Scatter}
\label{sec:scatter}

The relations in Figure~\ref{fig:S2016comp} are generally tight, with a scatter of 0.2 dex about the line of equality.
This scatter can be linked to contributions from physical effects (Section~\ref{sec:physdrive}) and observational effects (Section~\ref{sec:bandconv}).
In particular it is useful to understand their origin and potential for correction to achieve a tighter relationship.

Dust heated by active galactic nuclei (AGNs) is not explicitly included in our numerical experiment, but it is worth considering whether this can add additional scatter to the \Le-\Mmol relation.
We briefly consider the influence of this additional dust heating source in Section~\ref{sec:AGN}.

\subsection{Physical Properties of the Galaxies}
\label{sec:physdrive}

As noted, a \Le-dependent conversion factor from \Le to \Mmol seems to provide a slightly better fit for our simulated galaxies than a constant \Le-\Mmol relation, yet there is still scatter about the relation.
In order to investigate any physical origin of the scatter, we utilize the other galaxy properties available in the simulations and summarized in Figure~\ref{fig:simproperties}.
In Figure~\ref{fig:mmolcomp} we show the ratio of the \Le-inferred \Mmol values (using the \Le-dependent calibration) to the ``true'' \Mmol values, against $z$, \Tdust, \Mdust, the product $\langle {\rm T}_{\rm dust}\rangle_{\rm M}{\rm M}_{\rm dust}$ and \Mmol/\Mdust.

There are no clear trends of \Mmole/\Mmolt with redshift or \Tdust.
The latter is particularly interesting as it suggests that \Tdust-dependent corrections will not result in more accurate estimates of \Mmol from \Le.
Typically, the mass-weighted \Tdust is $\sim5-15$ K higher than the $25$ K assumed when doing band conversion, but this only affects the right column of Figure~\ref{fig:mmolcomp}.
Based on the left column of Figure~\ref{fig:mmolcomp}, which does not require any assumption about \Tdust, more accurate \Tdust measurements are unlikely to lead to more accurate \Mmol determinations.

In contrast, differences between \Mmole and \Mmolt show some correlation with \Mdust.
This indicates that at least some of the scatter in determining \Mmol relates to variations in the dust to gas ratio.
The product $\langle {\rm T}_{\rm dust}\rangle_{\rm M}{\rm M}_{\rm dust}$ (see Equation~\ref{eq:DGR}) shows similar correlation with \Mmole/\Mmolt as \Mdust alone does.

However, the tightest correlation is with the molecular gas to dust mass ratio (Figure~\ref{fig:mmolcomp} bottom row).
It is unclear if it is possible to correct for this trend a priori, though if metallicities are available these may be used to estimate the dust to gas ratio.
Encouragingly, the left and right columns of Figure~\ref{fig:mmolcomp} look similar, confirming that use of band conversion is not dominating the scatter in the relation.
However, careful inspection of the figure (particularly comparing the bottom row) shows that the band conversion does slightly increase the scatter in recovery of \Mmol.

\subsection{The Effect of Band Conversion}
\label{sec:bandconv}

The results obtained from the two techniques for determining \Le are qualitatively similar, though there are differences which arise from the band conversion.
Obtaining \Lei requires assumptions for values of $\beta$ and \Tdust, as well as the assumption that $\nu_{\rm rest}$ and $850~\mu $m lie on a single-temperature blackbody.
Here we discuss the origin of the differences between \Led and \Lei, in terms of SED complexity and potential mis-match between assumed and true parameters ($\beta$, \Tdust).

\subsubsection{Redshift-dependent Band Selection}

Here we discuss the impact of band conversion when observing bands are selected such that $\Gamma_{\rm RJ}$ is not too large.

\begin{figure*}
\centering
\includegraphics[width=0.45\textwidth]{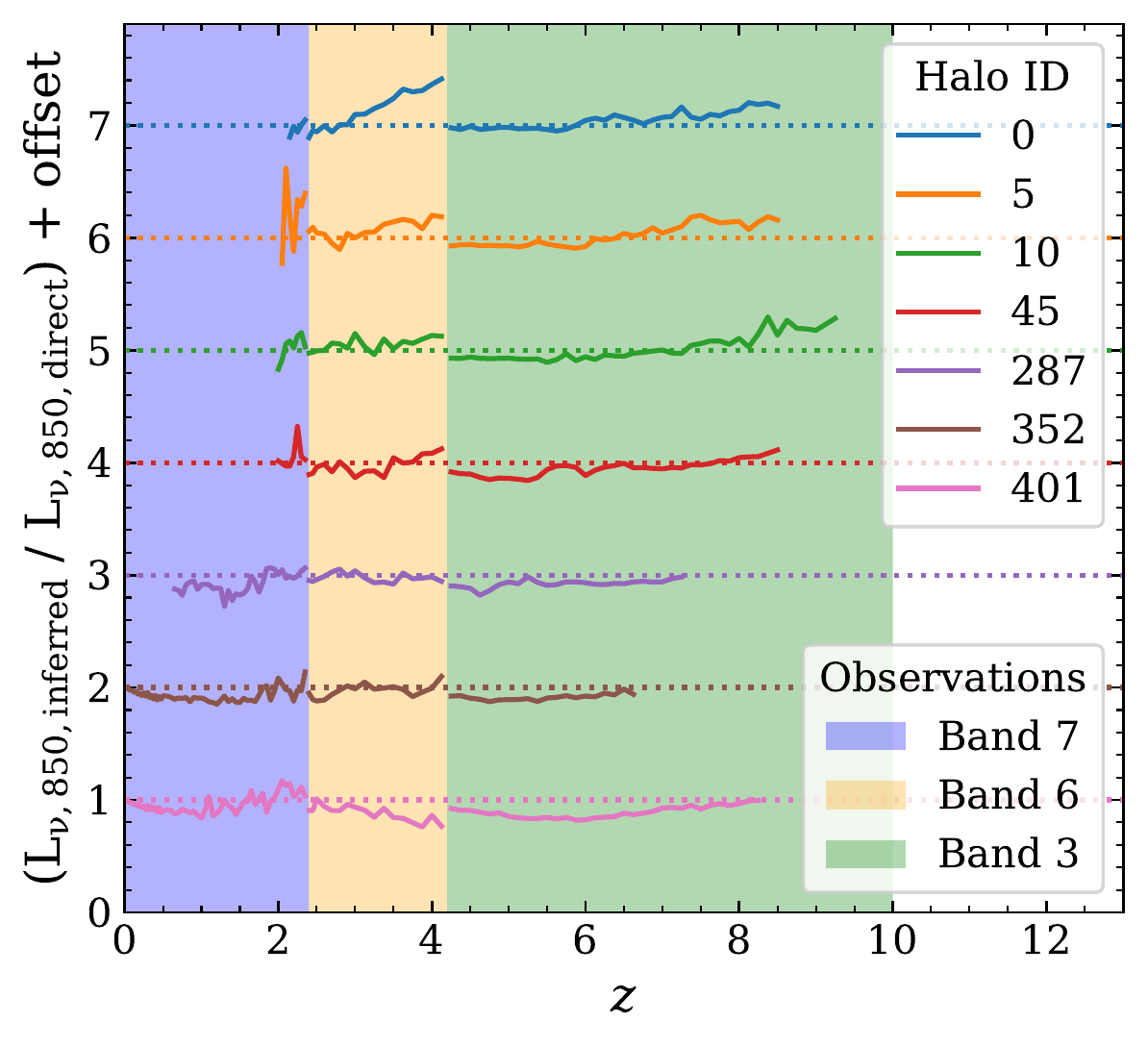}
\includegraphics[width=0.45\textwidth]{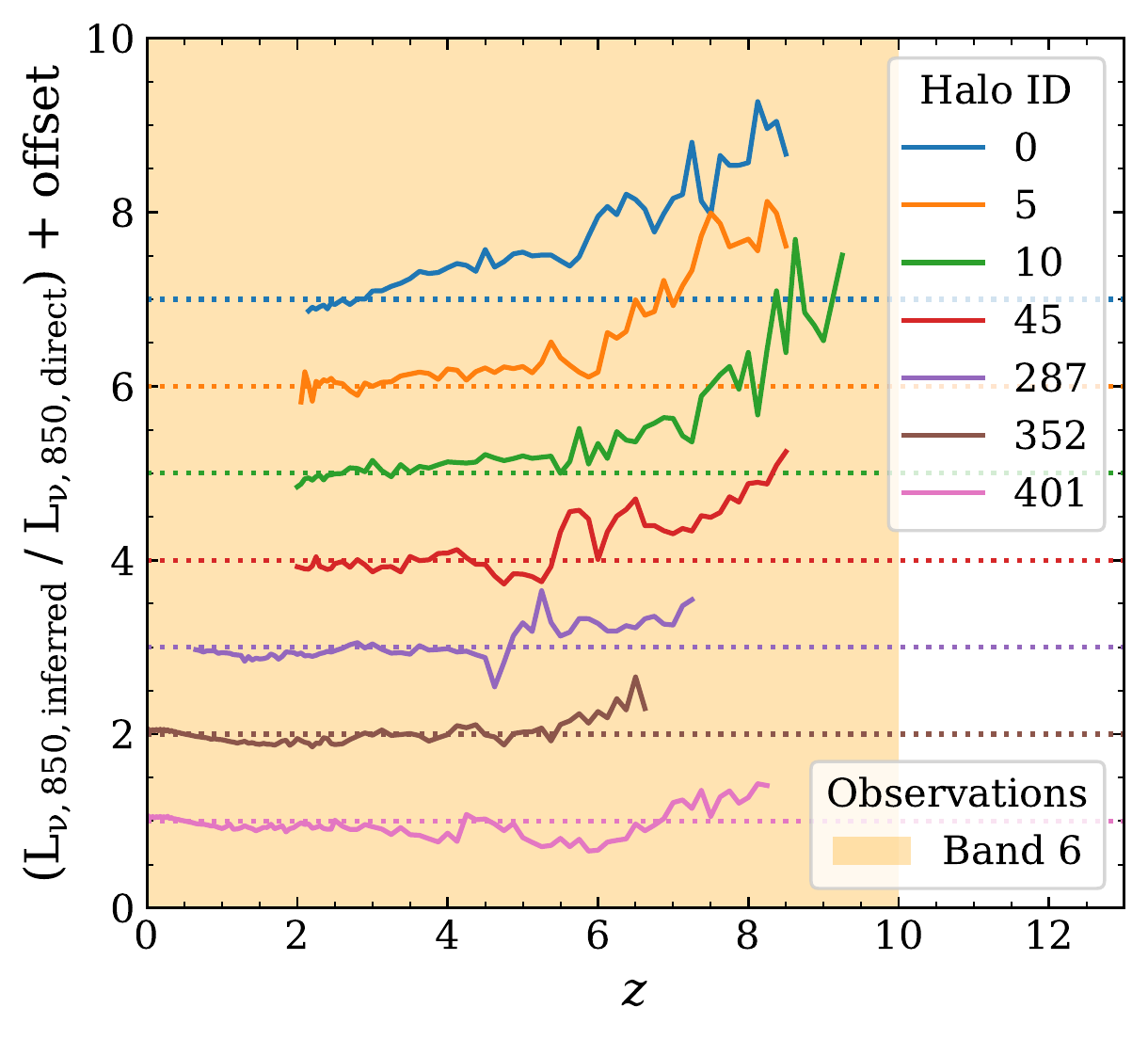}\\
\includegraphics[width=0.45\textwidth]{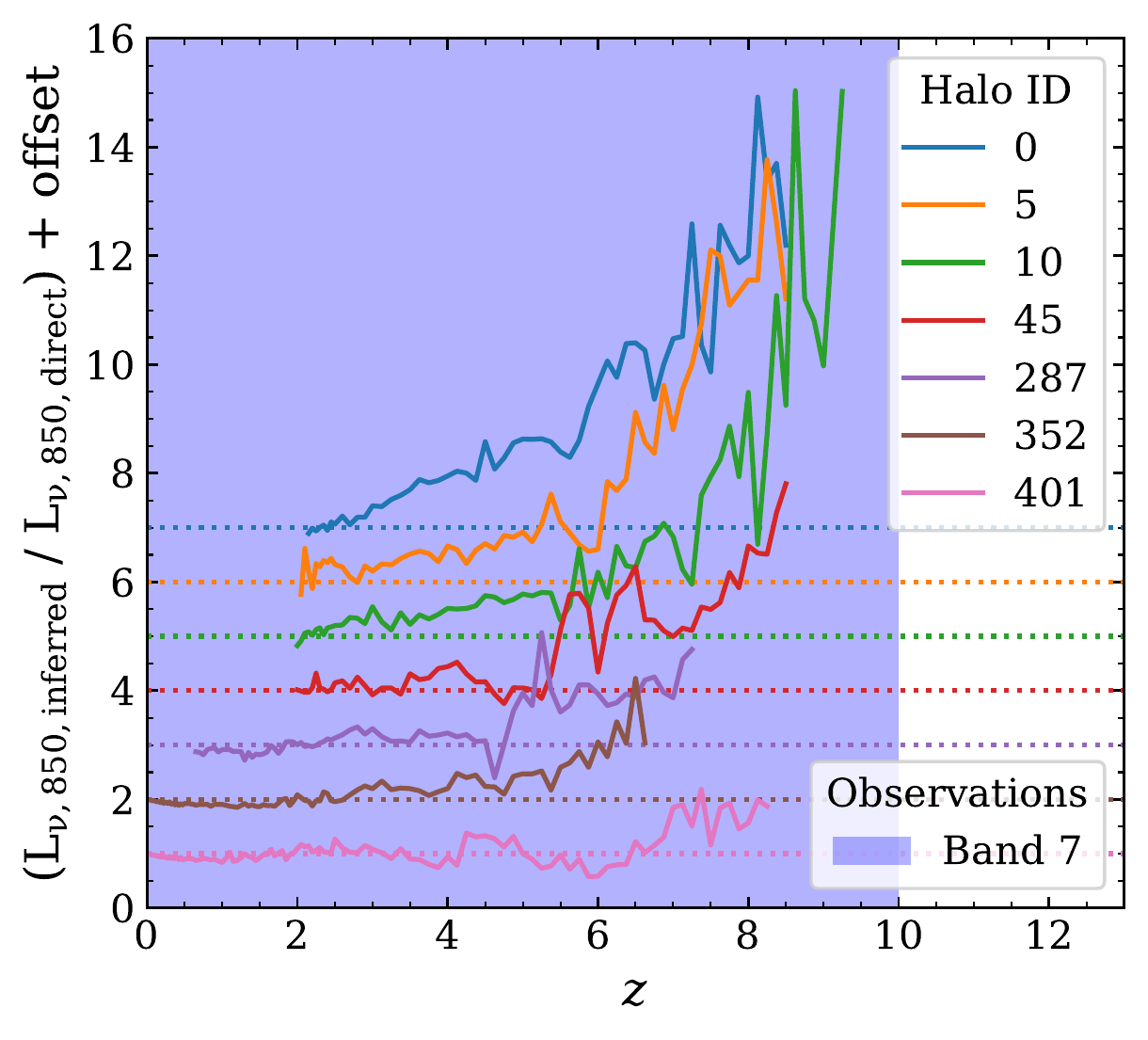}
\caption{The ratio of the \Lei inferred using a higher rest-frequency observation and down-converting to an equivalent $850~\mu m$ flux (Figure~\ref{fig:L850-z} right) to the directly measured \Led (Figure~\ref{fig:L850-z} left).
Clockwise from upper-left: a redshift-dependent band selection, Band 6 only, and Band 7 only.
Note the difference in y-scales between the panels.
For clarity the track for each halo is offset by a constant factor and the dotted lines mark the unity line for each individual halo.
In the upper left panel the discontinuities at $z=4.2$ and, to a lesser degree, at $z=2.4$ originate in the change in ALMA bands (see Appendix~\ref{app:bands}) and the fact that the dust SEDs are not single-temperature blackbodies.
Other deviations likely reflect mismatches between the assumed parameters (\Tdust, $\beta$) and their true quantities.
The use of ALMA Bands 6 and 7 at $z>4-6$ results in systematic overestimation of \Le, with discrepancies potentially exceeding 0.5 dex for the most massive halos.
}
\label{fig:L850-offset}
\end{figure*}

To more clearly show the differences in \Led and \Lei for the same snapshots, in Figure~\ref{fig:L850-offset} we show the ratio of \Lei to \Led.
The inferred \Lei varies with respect to the directly measured \Led, with variations typically on the order of $10-20\%$, but up to 50\%.
Broadly, the variations can be separated into two categories:
The first are sudden discontinuities at $z=4.2$ and $z=2.4$ which correspond to the change between ALMA bands.
The jumps reflect the fact that the SEDs are not perfect single-temperature blackbodies and may arise from mismatches in the assumed \Tdust and $\beta$, so the factor of 2.3 and 1.6 change in rest frequency suddenly probes a different portion of the SED.

The remainder of the variation likely reflects mis-matches between our assumed $\beta$ and \Tdust and the true values.
It is important to note that this is \emph{not} the uncertainty in the use of \Le as a mass estimator but instead reflects \emph{additional} uncertainty resulting from the process of inferring \Se from observations at a different rest frequency.
This can readily be seen in the bottom row of Figure~\ref{fig:mmolcomp} where the scatter is visibly increased for \Mmol estimates using \Lei, compared to estimates when \Led is known.
We further discuss these effects in more detail in Appendix~\ref{app:bands}.

In general the \Lei values track the directly measured \Led values when averaged over the evolution of the halo.
However in this sample of galaxy formation simulations, there are systematic features in Figure~\ref{fig:L850-offset} in certain redshift ranges for all halos.
The few snapshots at $z>8$ appear to have an inferred \Lei which systematically over-estimates \Led by $>10\%$.
This is likely due in part to \Tdust being $40-50$ K (Figure~\ref{fig:simproperties}) while we assume $25$ K in the band conversion.

Figure~\ref{fig:mmolcomp} shows the mass-weighted mean \Tdust is almost always larger than the 25 K we assume; this, coupled with our fiducial study of Appendix~\ref{app:bands} suggests underestimates of \Le from band conversion may result from the fact that we are sampling fluxes from a SED which is not a single-temperature blackbody.
It is unlikely this can be corrected without multi-band measurements.
However, this effect is small relative to the overall scatter in the \Le-\Mmol relation.

\subsubsection{Redshift-independent Band Selection}

Observing time constraints may necessitate dust continuum observations in a single band, independent of the source redshift.
In Figure~\ref{fig:L850-offset} we also show the ratio of \Lei/\Led as a function of redshift for our simulations, but assuming only Band 6 or Band 7 is used at all redshifts.
At high redshifts ($z\gtrsim5$) Bands 6 and 7 are sufficiently far from $850~\mu m$ that the recovery of \Le is significant compromised.
For the more massive halos at $z\sim8$, the discrepancy can exceed 0.5 dex.
In these cases \Lei is systematically above \Led, and so inferences of \Mmol would be biased high by the same amount (up to 0.5 dex).

\subsection{The Potential Impact of AGN-heated Dust}
\label{sec:AGN}

In these simulations and post-processing we have not included radiative contributions from active galactic nuclei (AGNs).
AGNs are known to have hot ($\sim1000$ K) dust in inner 10s of parsec, owing to the intense radiation fields.
Modeling of AGN tori suggests that hosts have dust warm dust masses of $10^3-10^5$ \Msun \citep{Fritz2006}.
These dust masses are significantly smaller than the typical dust masses of our simulated galaxies (Figure~\ref{fig:simproperties}, typically by factors $>10^3$.
In the optically thin limit, \Le is linearly dependent on \Tdust and \Mdust.
Though the \Tdust of AGN-heated dust can be a factor of $\sim30$ larger than the dust heated by star formation, the mass of dust at these high temperatures is $\sim 0.1\%$ of the total dust mass.
Combining these factors, the AGN-heated dust likely has a typical contribution to \Le on the order of a few percent.
This is much smaller than the scatter from physical processes (Section~\ref{sec:physdrive}) and observational techniques (Section~\ref{sec:bandconv}), and so is unlikely to be significant for the bulk of the galaxy population.
The class of extreme highly obscured quasars may have a significant component of their far-infrared emission generated via AGN heating of dust \citep[e.g.,][]{Wu2012,Tsai2015,Assef2015,Schneider2015,Diaz-Santos2016}.
Mid-infrared/submillimeter colors may be useful in identifying these objects which are likely to suffer from significant AGN contamination to the submillimeter \citep[e.g.,][]{Stanley2018}.

\section{Discussion}
\subsection{Implications for High-$z$}
\label{sec:ALMAobs}

\begin{figure*}
\centering
\includegraphics[width=0.45\textwidth]{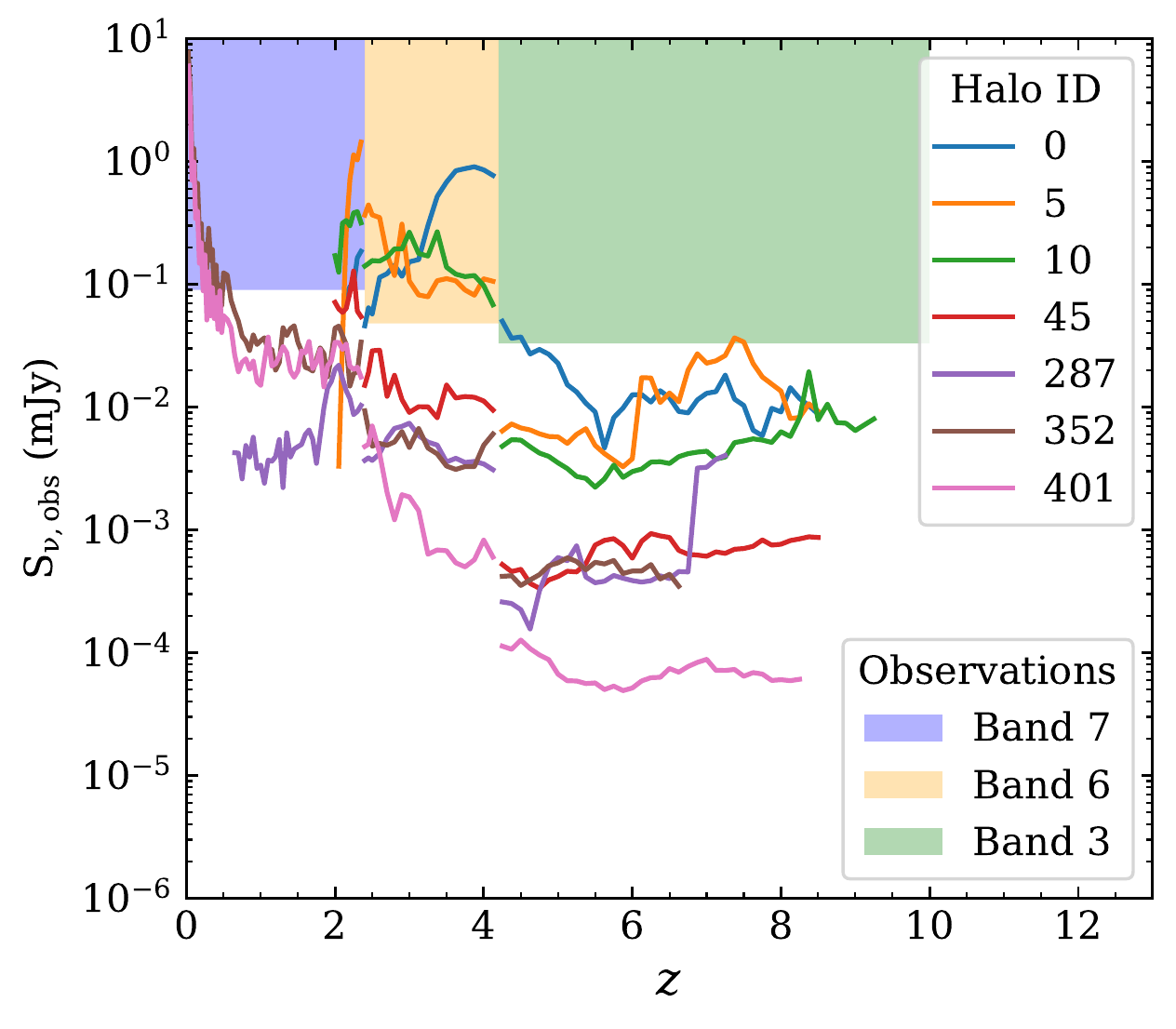}
\includegraphics[width=0.45\textwidth]{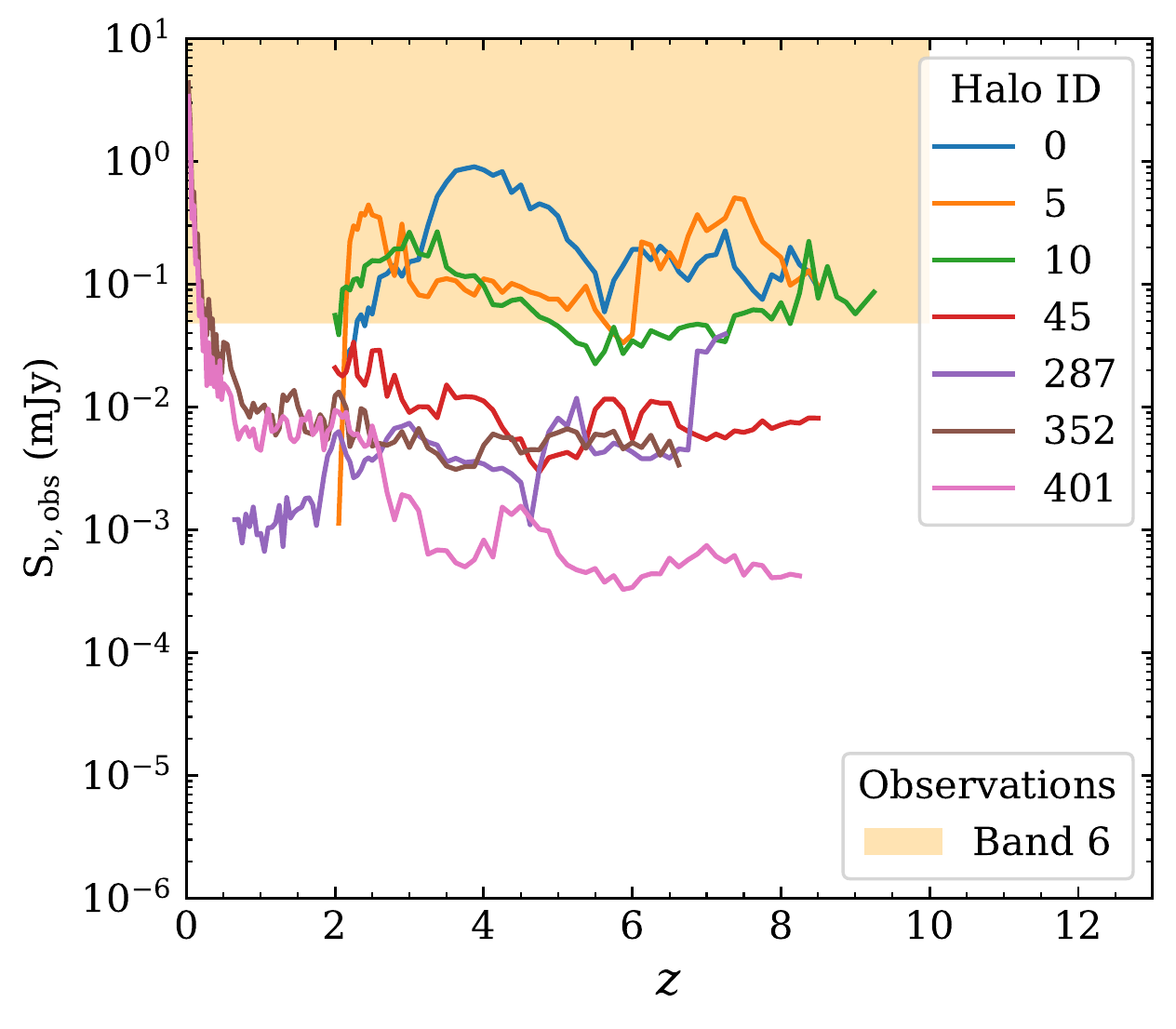}\\
\includegraphics[width=0.45\textwidth]{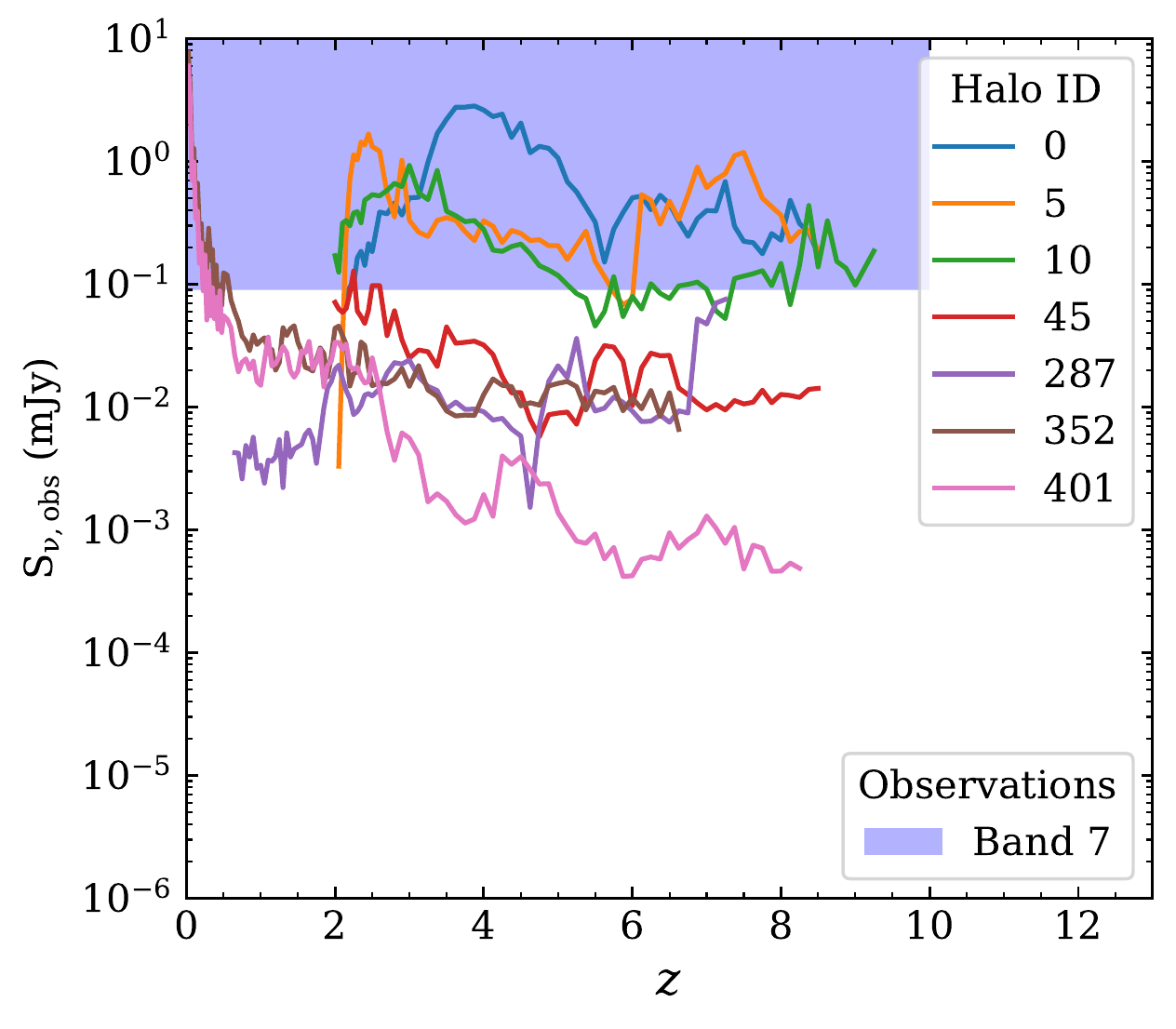}
\caption{The observed flux density as a function of redshift for the halos in the study, for the realistic case (Section~\ref{sec:redshifted}).
The lower edge of the colored boxes denote the $3\sigma$ detection limit for a one hour on-source ALMA observation in the noted band (i.e., halo tracks lying inside the colored boxes would likely be detected by ALMA in our fiducial 1 hour integrations).
Discontinuities in the flux curves for individual halos reflect the change in observing bands.
Clockwise from upper-left: a redshift-dependent band selection, Band 6 only, and Band 7 only.
The use of only Band 6 or Band 7 improves prospects for detection by taking advantage of the negative K-correction in the far infrared.
However, \Lei derived from these observations are systematically high, so \Mmol estimates will be similarly biased high (Section~\ref{sec:bandconv}).
These high-redshift detection prospects will be compromised if the dust-to-metals ratio is below the MW value (e.g., Figure~\ref{fig:RRdetect}).
}
\label{fig:detection}
\end{figure*}

We now briefly turn our attention to the approximate detectability of
these systems with ALMA.  In Figure~\ref{fig:detection} we show the
observed flux densities as a function of redshift, for the procedure
outlined in Section~\ref{sec:redshifted}, mimicking typical
observational approaches.  The tops of the colored boxes mark the
$3\sigma$ sensitivity of ALMA in one hour of on-source integration
(Table~\ref{table:ALMA}); i.e., tracks inside the boxes are detectable,
while tracks outside the boxes are not.

Over broad redshift ranges, most of our simulated galaxies are too faint to be detected in individual one hour observations. The most fertile range is around $z=3-4$ where the massive galaxies have high intrinsic dust luminosities.
At higher redshifts, the more massive galaxies/halos may be individual detectable with the aid of gravitational lensing \citep[e.g.,][]{Laporte2017}.

As we note in Appendix~\ref{app:CMB}, the inclusion of the CMB has an important effect on the intrinsic \Tdust (and hence, monochromatic luminosities) of the galaxies at high-redshift ($z\gtrsim6$)
The absolute detectability of such high-redshift systems will also depend on their contrast against the CMB \citep{daCunha2013b}, which we do not account for in Figure~\ref{fig:detection}.

\subsection{Dust Formation}
\label{sec:dustformation}

Large uncertainties currently exist in our understanding of the physics of dust formation.
Thus it is unclear how rapidly dust masses accumulate in the early universe.
Our fiducial model assumes the dust mass is a constant fraction of the metal mass.
However, there is observational evidence that the dust-to-metals ratio decreases at low metallicity \citep{Remy-Ruyer2014}, possibly connected to lower efficiencies of dust production in the ISM \citep{Zhukovska2014,Popping2017} or outflows \citep{Feldmann2015}.
This may lead to a further decrease in \Le at high-redshifts if the dust is not produced rapidly enough or if it is removed \citep[but see][for evidence of rapid dust production by $z \approx 7-8$]{Laporte2017,strandet17a,marrone18a}.
This would presumably further increase the molecular gas to dust mass ratio, resulting in more significant underestimates of \Mmol.

\begin{figure}
\centering
\includegraphics[width=0.45\textwidth]{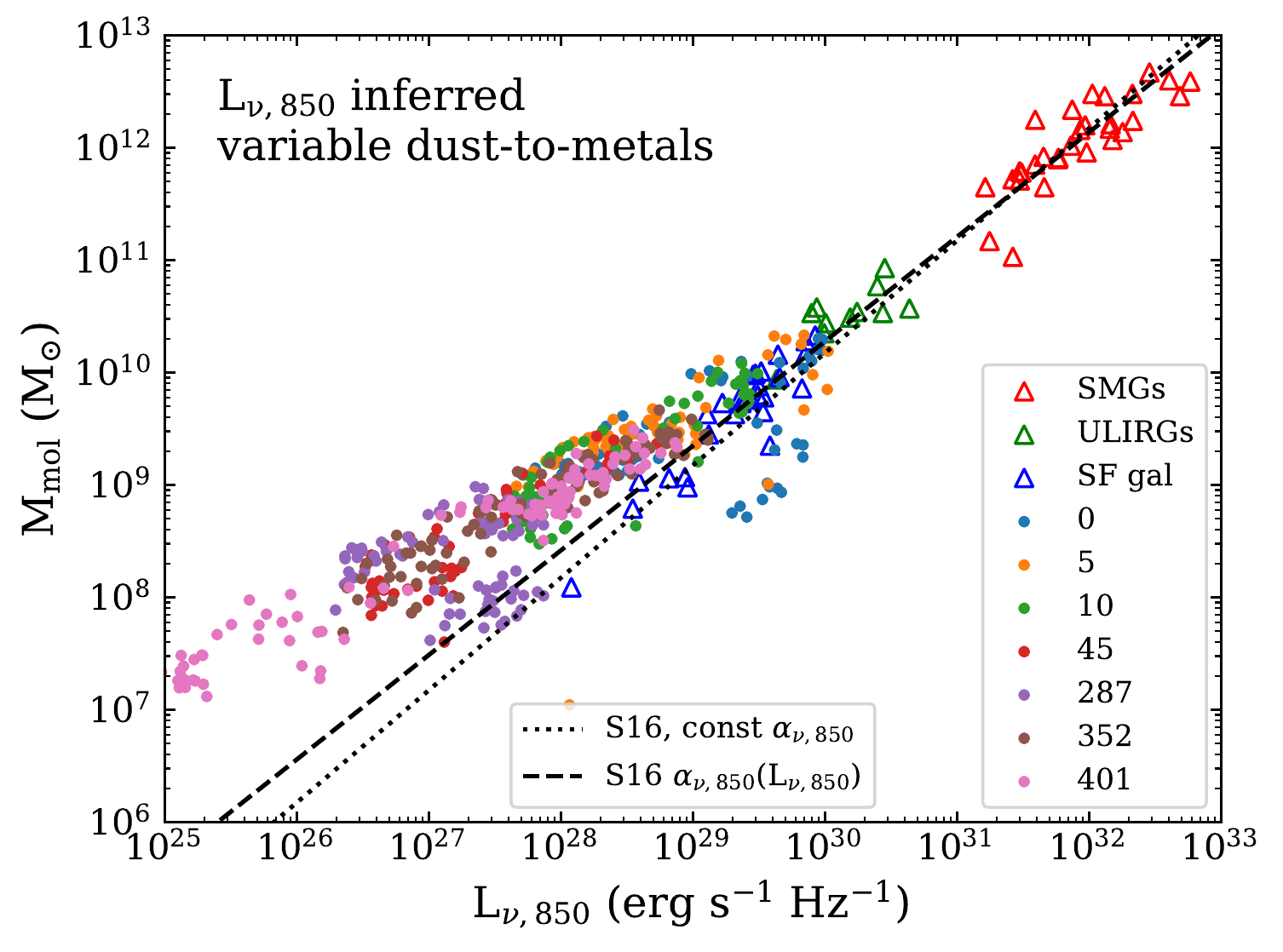}
\includegraphics[width=0.45\textwidth]{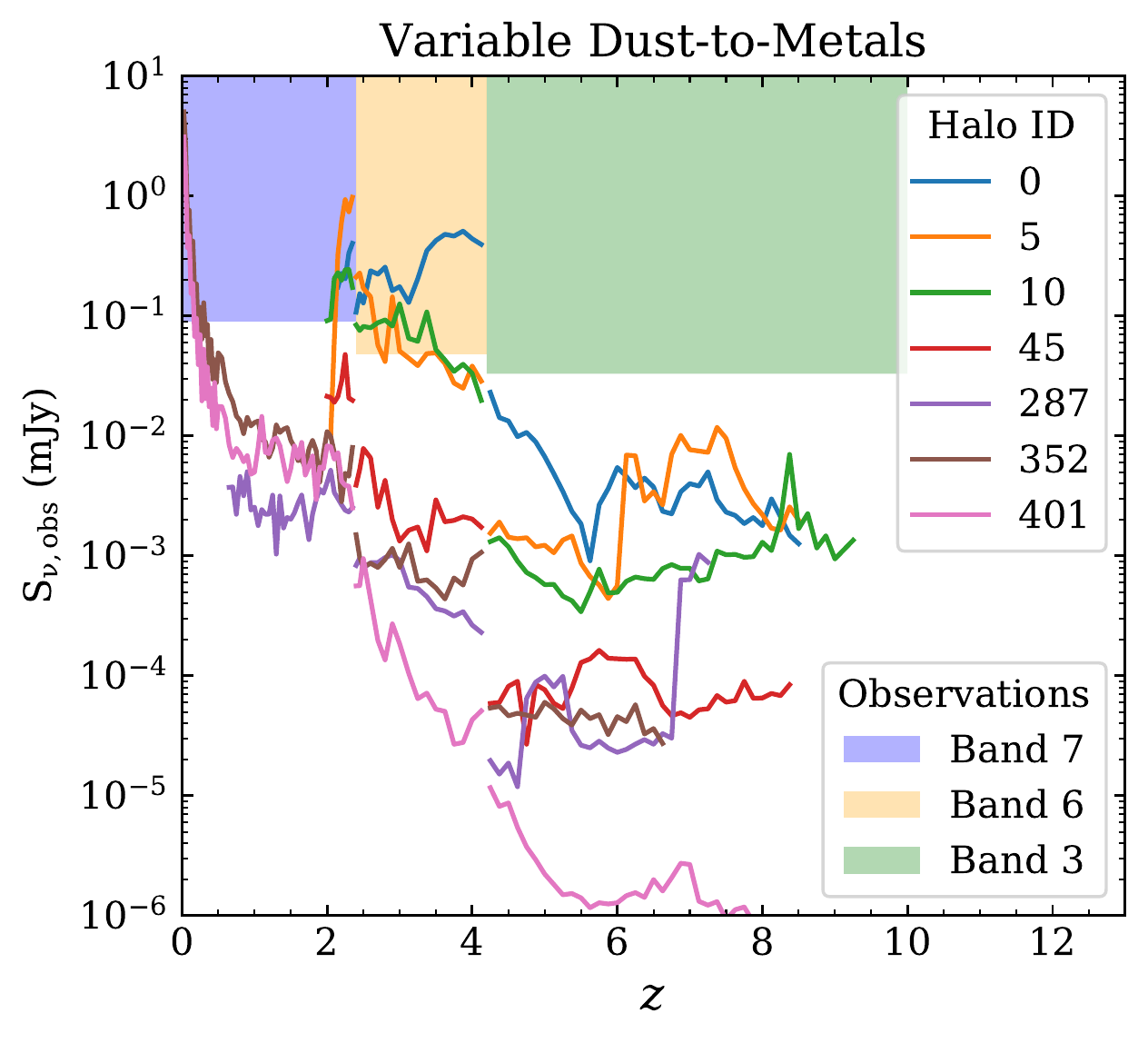}
\caption{Top: The molecular mass in the simulation as a function of \Le for the metallicity-dependent dust-to-metals ratio (compare to Figure~\ref{fig:S2016comp}).
Here the points are color-coded by the mass-weighted mean metallicity inside the 50 kpc box.
Bottom: the observed flux density for the halos as a function of redshift for the same model (compare to Figure~\ref{fig:detection}).
This dust model predicts a \Le-\Mmol relation in which \Le has a steep dependence on \Mmol.
The scatter about the trend also increases.
This is due to the variable dust-to-metals enhancing the effect of the gas within a galaxy having a range of gas metallicities.
The reduced dust masses also make these galaxies more difficult to detect.
}
\label{fig:RRdetect}
\end{figure}

To explore the qualitative behavior we implemented a variable dust-to-metals model, following the observational results of \citet{Remy-Ruyer2014}.
The details of this implementation and its effects are discussed in Appendix~\ref{app:RR}. In short, we computed a metallicity-dependent dust-to-metals ratio for each gas element.
This second-order effect has a noticeable impact on both the relation between \Le and \Mmol and the implied detectability (Figure~\ref{fig:RRdetect}).
\Le has a much steeper dependence on \Mmol (i.e., on average \Le increases significantly with small increases in \Mmol).
Additionally, the scatter about the trend increases, reflecting the presence of a range of dust-to-metals ratios within a single halo.
The net effect is that detecting continuum emission in lower-mass / lower-metallicity galaxies will be more difficult than may be expected from the \citet{Scoville2016} relation and the inferred molecular masses will be more uncertain.
Comparing Figure~\ref{fig:RRdetect} (top) with Figure~\ref{fig:S2016comp} suggests that a variable dust-to-metals ratio implies galaxies may deviate significantly ($\gtrsim0.5$ dex) from the \citet{Scoville2016} relation below luminosities of \Le$\lesssim10^{28}$ ergs s$^{-1}$.
However, this low-metallicity regime is similar to where simulation \HH masses become sensitive to the adopted formation model \citep[e.g.,][]{Popping2014,Lagos2015}, meaning a more precise calibration would need to consider the effects of varying \HH formation.
In the future, models that incorporate the self-consistent formation and destruction of dust in galaxy formation simulations will be necessary to fully understand the uncertainties involved in deriving \Mmol from \Led \citep[e.g.][Q. Li et al. \emph{in preparation}]{mckinnon16a,Popping2017}.

\subsection{Comparison with Other Simulations}

The FIRE collaboration has also explored the \Le-\Mmol relation using their suite of cosmological zoom simulations.
\citet{liang18a} explored this relation between $z=2-4$ for galaxies with M$_{*}\gtrsim 10^{10}$ \Msun.
Within that redshift range, our halos 0, 5, and 10 overlap with the stellar mass range of their simulations (see Figure~\ref{fig:simproperties}).
Considering those halos within that restricted redshift range and our fiducial model for the dust-to-metals ratio, we find good agreement with their results that \Le can accurately recover \Mmol and that the scatter is primarily dependent on the gas to dust ratio.
This agreement is encouraging: the underlying physics driving the evolution of the \citet{liang18a} galaxies is substantially different than those governing our own models.
Beyond this, Liang et al. include the effect of obscuration by birth clouds around young stars.
The general agreement between these results and our own further supports the robustness of the \Le-\Mmol relation for massive galaxies in this redshift range.

\section{Conclusions}

We present analysis of the $850~\mu m$ emission in simulated SEDs derived from radiative transfer post-processing of hydrodynamic cosmological zoom simulations.
We find:

\begin{itemize}
    \item{ \Le correlates well with \Mmol in our simulations, confirming the viability of using the 850~$\mu$m emission as a molecular mass tracer for massive galaxies, independent of the assumed dust model.
    We find the best agreement with the \Mmol values in the simulation by employing the \Le-dependent calibration factor of \citet{Scoville2016}.
    Despite the fact that the \Le-dependent and constant calibrations are typically within a factor of two, the difference is systematic so we recommend the use of \Le-dependent variation.}
      
    \item{The band conversion of redshifted flux measurements to rest-frame $850~\mu m$ fluxes can introduce errors typically on the order of $\sim10-20\%$ (but sometimes up to 50\%). These errors arise from mis-matches in the parameters used in the band conversion (\Tdust, $\beta$) as well as from the fact that the SEDs are not single-temperature blackbodies.}
    \item{The scatter in the relation appears to be set primarily by the variations in the molecular gas to dust mass ratio.
    Despite a $\sim30$ K range in \Tdust within our simulations, offsets from the mean \Le-\Mmol relation were not correlated with \Tdust.}
    \item{Exploration of a variable dust-to-metals prescription suggests lower-metallicity / lower-mass galaxies may deviate from published \Le-\Mmol relations and show increased scatter about those relations.
    Deviations may become significant (i.e., $\gtrsim0.5$ dex) below \Le$\lesssim10^{28}$ erg s$^{-1}$ or $\log_{10}({\rm Z}/{\rm Z}_{\odot}) \lesssim -0.8$.
    However, detailed dust formation models in cosmological simulations are needed to make specific predictions.}
\end{itemize}

\acknowledgements

The authors thank the anonymous referee for his/her comments, which have improved the quality of the paper.
The authors thank L. Armus, C. Cicone, A. Gowardhan, R. Feldmann, and L. Liang for comments on an earlier version of this paper.
The authors thank M. Stalevski for information on the dust masses of AGN hosts and thank S. Linden for information on the clearing timescales for young star clusters.
G.C.P. acknowledges support from the University of Florida and thanks the Sexten Center for Astrophysics (\url{http://www.sexten-cfa.eu}) where part of this work was performed.
D.N. was funded in part by grants NSF AST-1715206 and HST AR-15043.0001.
A portion of this work was performed at the Aspen Center for Physics, which is supported by National Science Foundation grant PHY-1607611.

This research has made use of the NASA/IPAC Extragalactic Database (NED) which is operated by the Jet Propulsion Laboratory, California Institute of Technology, under contract with the National Aeronautics and Space Administration.
This research has made use of NASA's Astrophysics Data System.
The authors are grateful to the University of Florida Research Computing for providing computational resources and support that have contributed to the research results reported in this publication (\url{http://researchcomputing.ufl.edu}).

\software{ipython \citep{Perez2007},
numpy \citep{Vanderwalt2011},
matplotlib \citep{Hunter2007},
Astropy \citep{astropy,Astropy2},
the \texttt{dust\_emissivity} package (\url{https://github.com/keflavich/dust_emissivity}),
{\sc yt} \citep{turk11a},
{\sc hyperion} \citep{robitaille11a,robitaille12a},
{\sc fsps} \citep{conroy09b,conroy10b,conroy10a},
and
{\sc powderday} \citep{narayanan15a,Narayanan2018b}.}

\bibliography{ms,narayanan_bib}
\clearpage
\appendix

\section{Mismatches in Assumed \Tdust and $\beta$}
\label{app:bands}

\begin{figure}[h]
\includegraphics[width=0.5\textwidth]{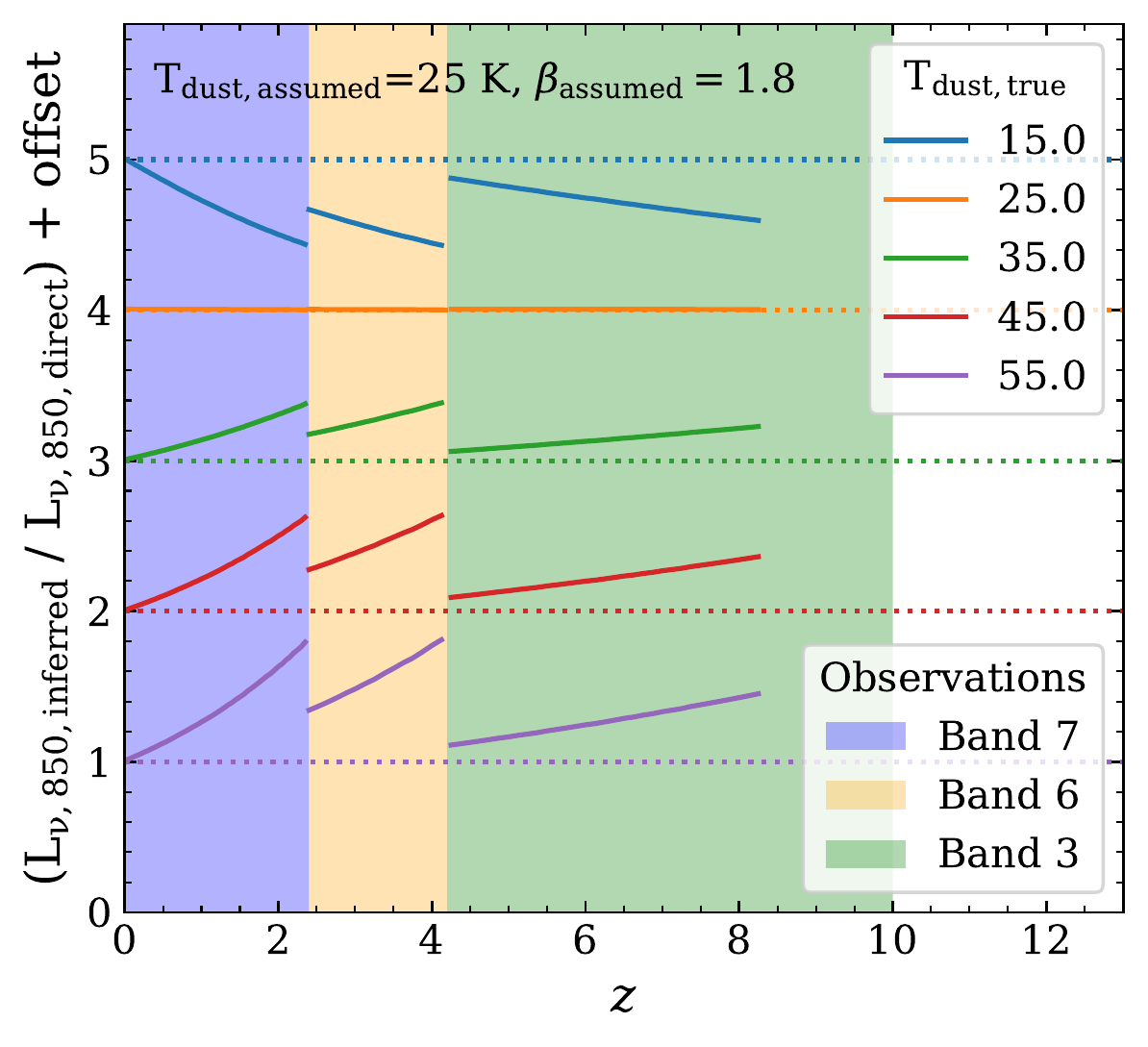}
\includegraphics[width=0.5\textwidth]{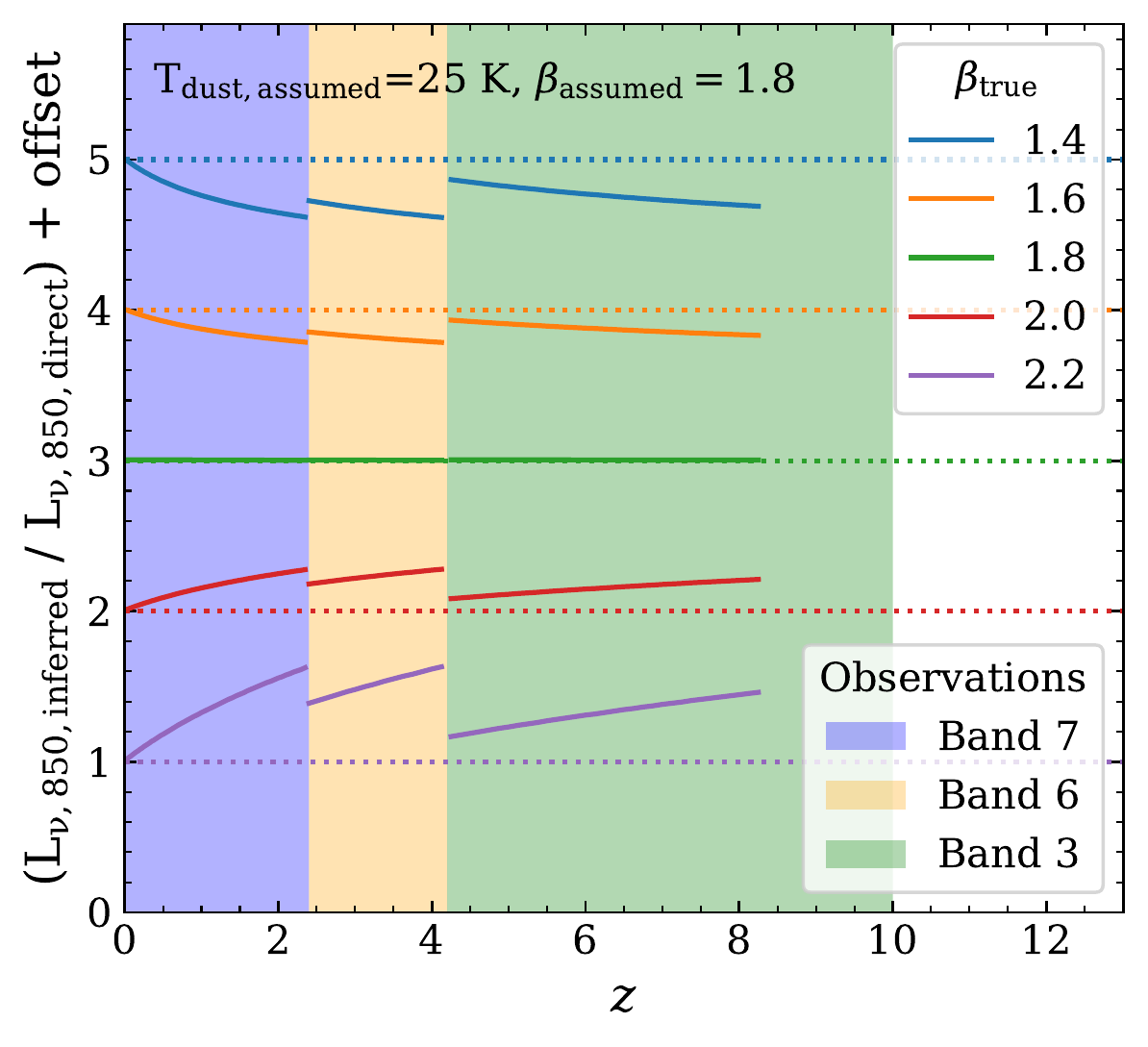}
\caption{
The application of the band conversion method of obtaining \Lei (Section~\ref{sec:redshifted}) applied to ideal single-temperature blackbodies.
Both panels show the ratio of \Lei/\Led as a function of redshift, employing the same ALMA band selection criterion used for the simulations in this paper.
Left: We vary \Tdust of the blackbody in the range $15-55$ K (with $\beta$ fixed to 1.8) and examine the accuracy of \Lei when we assume 25 K and $\beta=1.8$ for the band conversion.
Right: Holding \Tdust fixed at 25 K, we vary $\beta$ between $1.4-2.2$ and examine the accuracy of \Lei when we assume 25 K and $\beta=1.8$ for the band conversion.
Underestimating \Tdust or $\beta$ results in an underestimation of \Le when performing the band conversion.
For single-temperature modified blackbodies, the discrepancies can be $>50\%$ if the mis-matches are significant.
}
\label{fig:app:bbtest}
\end{figure}

Here we explore some of the effects of mis-matches between the assumed and true \Tdust and $\beta$ values.
In Figure~\ref{fig:app:bbtest} we show the ratio of the inferred \Lei to the true \Led for single-temperature modified blackbodies with a range of temperatures, redshifted and ``observed'' following the procedure in Section~\ref{sec:redshifted} and assuming \Tdust$=25$ K and $\beta=1.8$ for the band conversion.
The two panels explore mismatches in \Tdust (left) and $\beta$ (right), compared to our assumed values.
In both cases, underestimating (overestimating) the value of a parameter leads to an underestimation (overestimation) of \Le.
This can be thought of as a manifestation of the know T--$\beta$ degeneracy in SED fitting, where a high value of one parameter can be compensated for with a lower value of the other parameter.
Figure~\ref{fig:app:bbtest} clearly demonstrates that the jumps at $z=2.4$ and $z=4.2$ (corresponding to abrupt changes in the ALMA bands) reflect the impact of mis-matched T and/or $\beta$ assumptions, even in the ideal case of a single-temperature blackbody.
Note that the exact locations and magnitudes of these jumps depend on the exact choice of frequencies with the ALMA bands, the redshift ranges over which each band is used to infer \Le, and the degree of the mismatch in band conversion parameters (\Tdust, $\beta$).

\begin{figure}[h]
\centering
\includegraphics[width=0.45\textwidth]{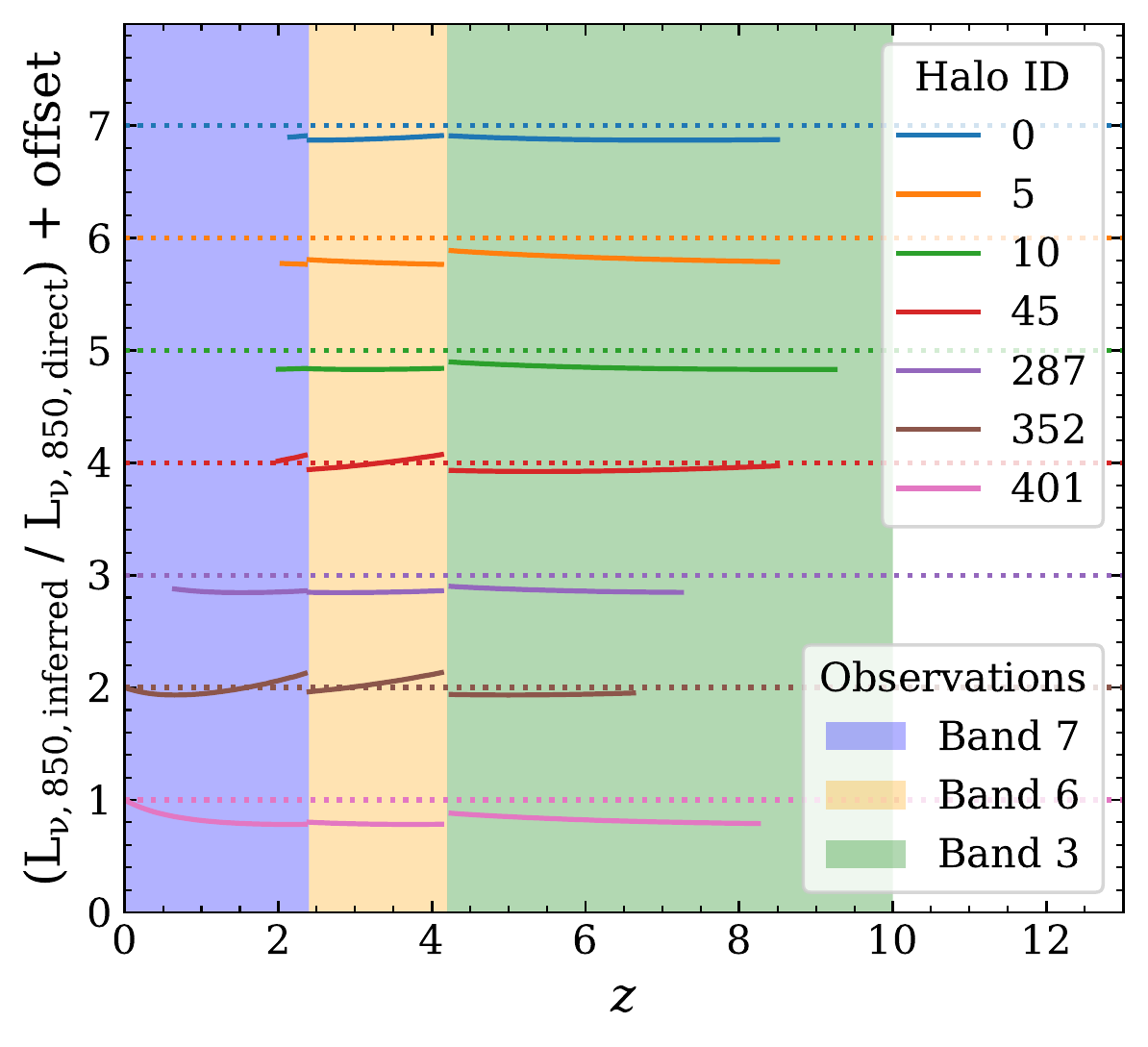}
\includegraphics[width=0.45\textwidth]{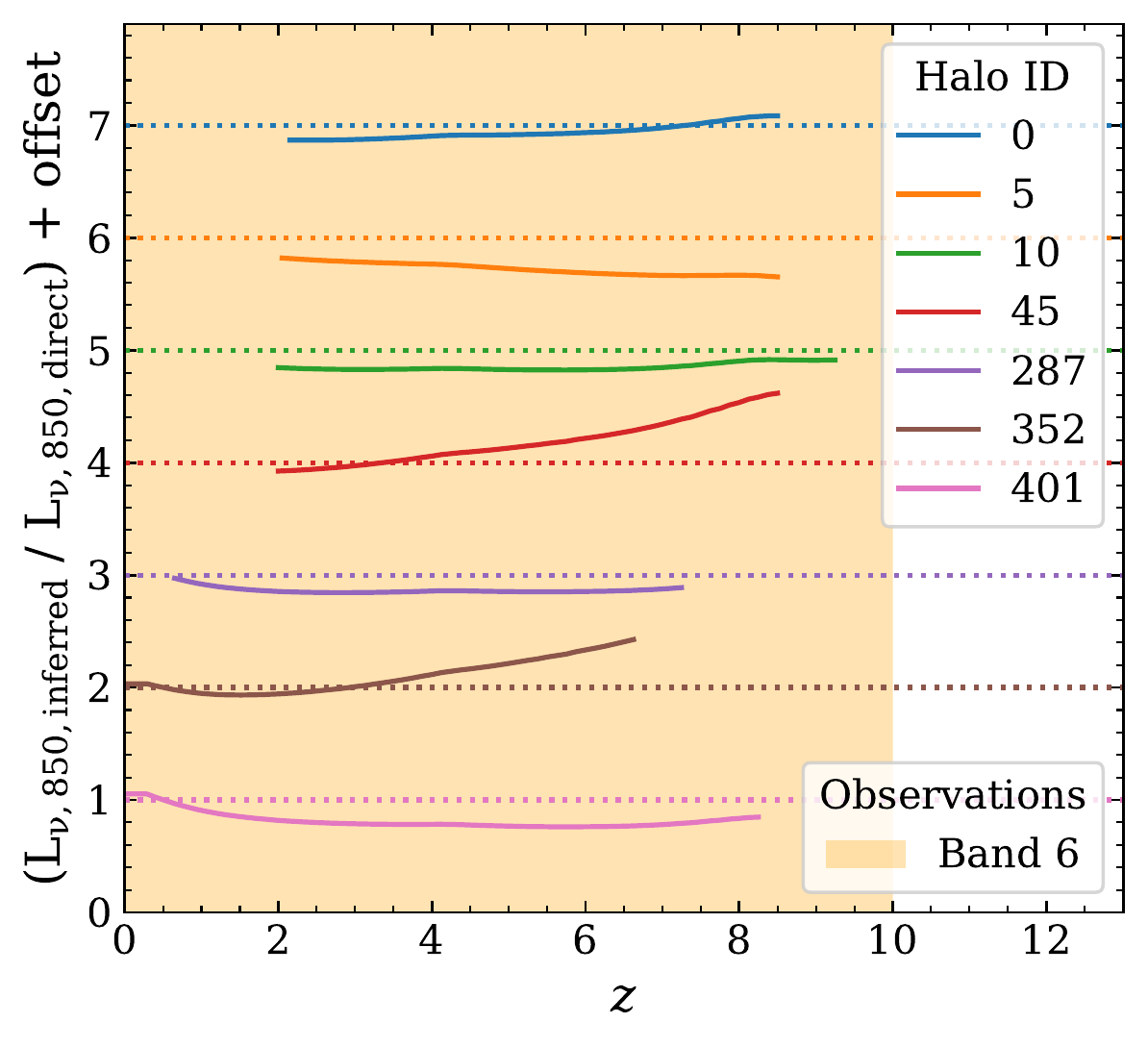}\\
\includegraphics[width=0.45\textwidth]{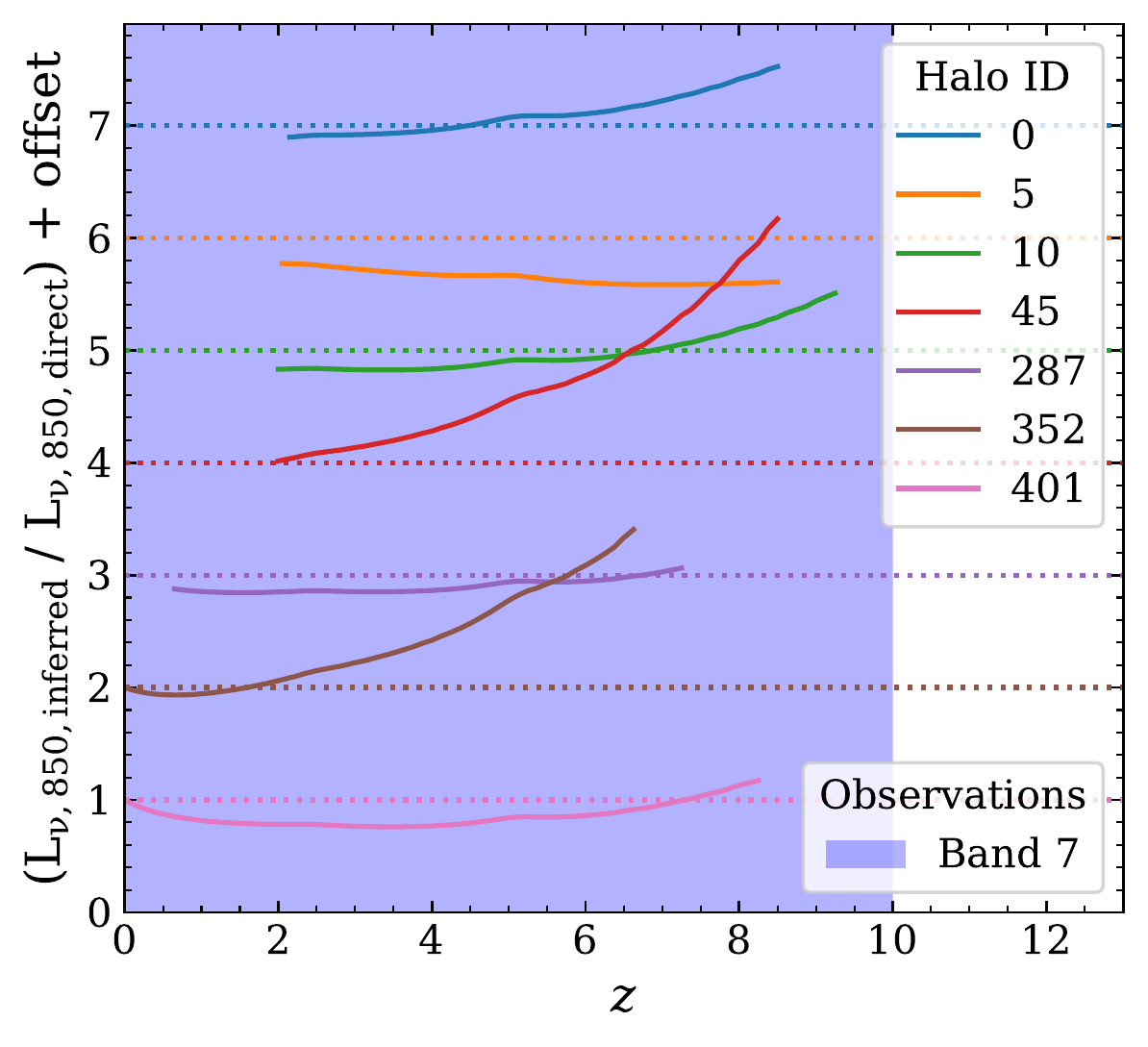}
\caption{Here we perform an analogous study to that in Figure~\ref{fig:app:bbtest}, but instead of using an ideal modified blackbody, we use an SED from the zoom simulations.
We take the SED of the snapshot at $z_{\rm final}$ for each halo (Table~\ref{table:simsum}, redshift it across the range probed by that halo and evaluate \Lei and \Led.
This shows how the band changes affect the accuracy of \Lei determination for realistic SEDs, while removing the effects of source evolution (e.g., in \Tdust).
Clockwise from upper-left: a redshift-dependent band selection, Band 6 only, and Band 7 only.
Comparing to the ideal blackbodies (Figure~\ref{fig:app:bbtest}), \Lei is consistently underestimated for most halos when using redshift-dependent observing bands, suggesting our band conversion is overestimating \Tdust and/or $\beta$.
Note that the jumps at $z=2.4$ and $z=4.4$ are still visible with a redshift-dependent band selection.
}
\label{fig:app:singsnap}
\end{figure}

These effects are evident when using more realistic SEDs from the cosmological zoom simulations.
In Figure~\ref{fig:app:singsnap} we show the ratio of inferred to true \Le for SEDs generated with {\sc powderday}.
Unlike in Figure~\ref{fig:L850-z} (right) where the intrinsic SED evolves with redshift, here we select the lowest-redshift SED for each halo and explore what \Lei would be inferred for a range of redshifts.
In essence, this is a somewhat more realistic version of the test in Figure~\ref{fig:app:bbtest}, using an SED which is not a single-component blackbody.
The sudden jumps are still present as in Figure~\ref{fig:app:bbtest}.

Mis-matches in \Tdust and $\beta$ can be up to $\sim80\%$ (e.g., Figure~\ref{fig:app:bbtest}) if the true \Tdust is $\sim2\times$ higher than we have assumed or if $\beta$ is $\sim20\%$ larger than we have assumed.
However the single-snapshot test (Figure~\ref{fig:app:singsnap}) suggests that the SEDs are typically not this pathological and that \Tdust and $\beta$ mis-matches cause errors on the order of $20\%$ if observing bands are selected on the basis of the source redshift.
This is consistent with the increased scatter seen in Figure~\ref{fig:mmolcomp} (right).

In scenarios where ALMA Band 6 or 7 is used for sources at all redshifts, the discrepancy is magnified.
Because \Tdust is not changing in this simplified exploration, this suggests that the use of rest frequencies far from $850~\mu m$ exacerbates the effect of multiple temperature components in the SEDs.

The question remains of how to determine the value of \Tdust to use when performing the band conversion.
As \citet{Scoville2014} noted, SED fitting results in an estimate for the luminosity-weighted \Tdust but it is unclear how discrepant this is with the mass-weighted \Tdust.
In Paper II we will more closely examine the link between \Tdust values inferred from SEDs and the true underlying distribution of \Tdust.

\section{Inclusion of the CMB}
\label{app:CMB}

The CMB effectively imparts a temperature floor on a galaxy.
Though generally negligible at low to intermediate redshifts, at high redshift the CMB approaches the rough dust temperatures of galaxies and so needs to be considered as a heating term.
This heating effect has been described by \citet{daCunha2013b} for ideal blackbodies, but here we briefly discuss the effect on simulated galaxies.

\begin{figure}
\includegraphics[width=\textwidth]{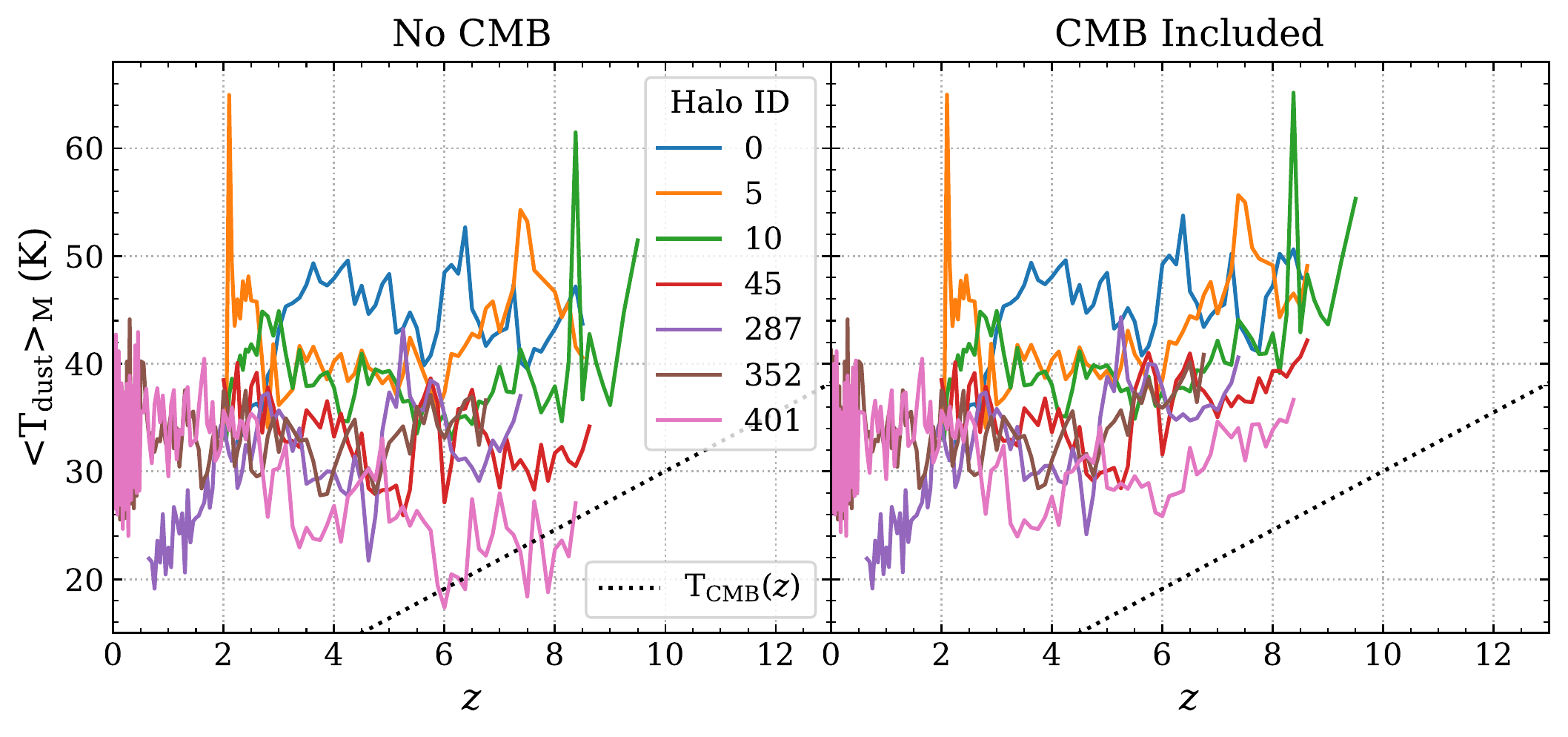}
\caption{Comparison of the mass weighted \Tdust derived from the {\sc powderday} processing without (left) and with (right) the CMB.
In both panels the dotted line shows the temperature of the CMB as a function of redshift.
The inclusion of the CMB affects \Tdust values for $z\gtrsim 6$, with the most pronounced differences originating in the lowest mass halos.
In extreme cases, the predicted \Tdust values can fall below the CMB temperature if the CMB is not included.}
\label{fig:app:CMB}
\end{figure}

Inclusion of the CMB in the radiative transfer postprocessing of the simulations has a noticeable effect on the mass-weighted mean \Tdust values for halos at $z\gtrsim5$ (Figure~\ref{fig:app:CMB}), with increases of $10-50\%$, depending on the halo mass and redshift.
This in turn translates to a higher \Le (Equation~\ref{eq:DGR}).
Based on the agreement between the {\sc powderday} \Le (including the CMB) and the simulation \Mmol (Figures~\ref{fig:S2016comp} and \ref{fig:mmolcomp}) accounting for the CMB heating is important for recovering the \Le-\Mmol relation for high-redshift galaxies.
The increased \Tdust values only partly compensate for the building dust masses.
We remind the reader that contrast effects against the CMB may complicate the measurement of \Le \citep{daCunha2013b}.

\section{Variable Dust-to-Metals}
\label{app:RR}

In order to explore the impact of variable dust-to-metals ratios on \Le and $\alpha_{\nu,850}$ we implemented a metallicity-dependent dust-to-gas ratio in {\sc powderday}.
\citet{Remy-Ruyer2014} parameterized this in terms of the gas-to-dust ratio as a function of $12+\log_{10}({\rm O}/{\rm H})$.
We followed their fit of a single-powerlaw, using a metallicity-dependent CO to \HH conversion factor \citep[Table 1 of][]{Remy-Ruyer2014}:

\begin{equation}
\log_{10}({\rm GDR}) = 2.21 + 2.02(8.69 - 12 + \log_{10}({\rm O}/{\rm H}))
\end{equation}

We note that \citet{Remy-Ruyer2014} advocate for the double powerlaw relation and this is also supported by some models \citep[e.g.,][]{Asano2013,Zhukovska2014,Feldmann2015,Popping2017}.
These are cast in terms of the galaxy metallicity while our sub-grid model necessarily refers to the metallicity of individual gas particles.
It is unclear how to translate between these two, so we adopt a model which does not have a discrete transition and broadly reproduces the behavior of the observations.
This enables us to broadly explore the impact of variable dust-to-metals in the simulations.
We expect that adopting the double powerlaw model would result in galaxies with metallicities higher than the break metallicity lying on the observed \Le-\Mmol relation, while those with lower metallicites would deviate from the relation.
This deviation at low metallicities could be stronger for the lowest metallicity halos, owing to the steeper gas-to-dust ratio versus metallicity relation for the double powerlaw model.
However, future simulations explicitly treating dust formation and destruction (Q. Li et al. \emph{in preparation}) will negate the need for such coarse sub-grid prescriptions.

A variable dust-to-metals ratio is a second-order effect on the dust-to-gas ratio -- decreasing metallicity reduces the dust mass in two ways: an overall reduced reservoir of metals and a smaller fraction of that reduced reservoir is in dust.
We apply this prescription to individual gas elements within our snapshots, with the result that the gas in every snapshot has a range in the dust-to-metals ratio.

\begin{figure}
\includegraphics[width=\textwidth]{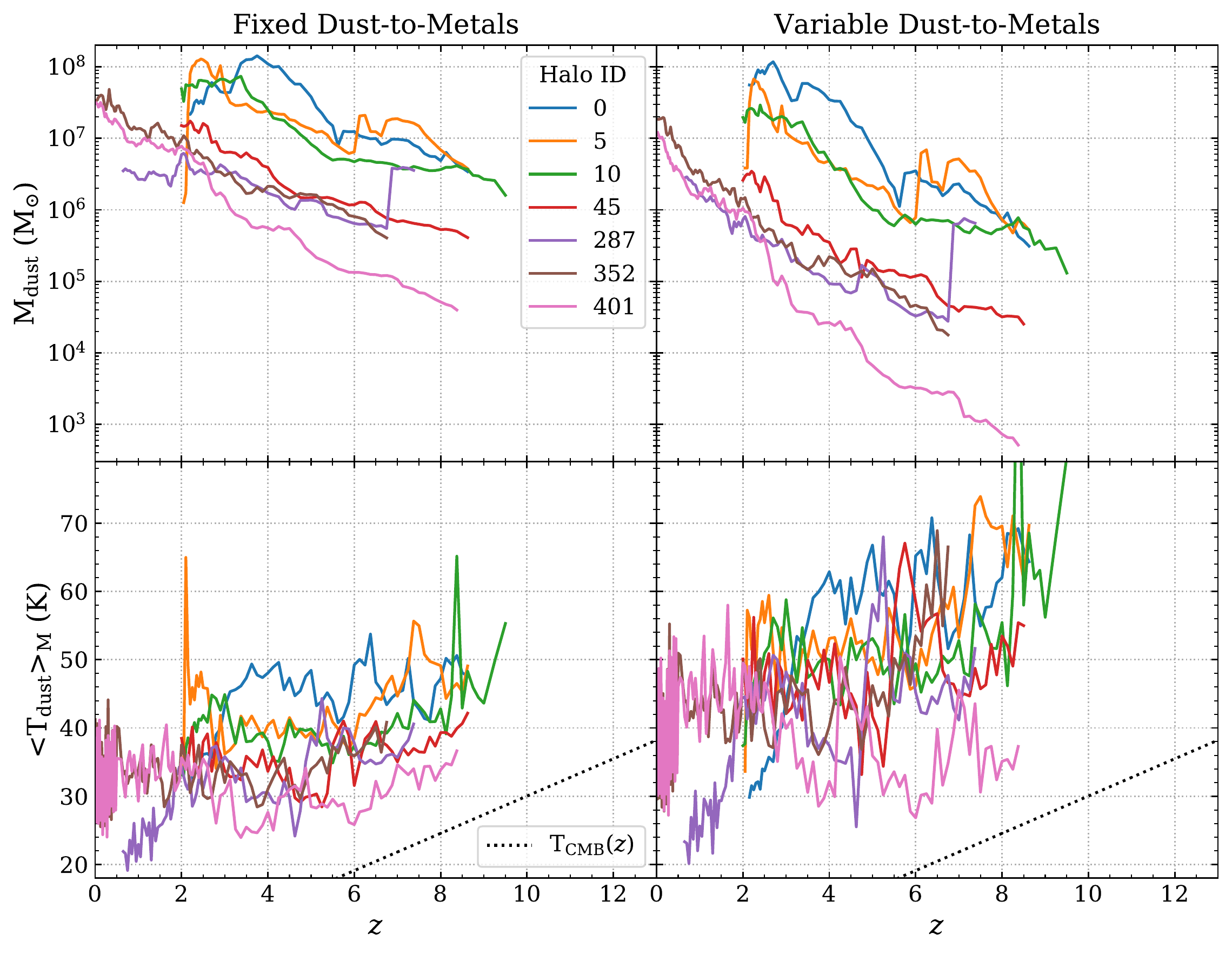}
\caption{The impact of the \citet{Remy-Ruyer2014} dust prescription on the inferred dust masses and temperatures in the zoom simulations.
Left: the fiducial case where the dust is a fixed fraction of the metal mass (identical to Figure~\ref{fig:simproperties}).
Right: a metallicity-dependent gas-to-dust ratio following the \citet{Remy-Ruyer2014} prescription described in the text.
The metallicity-dependent gas-to-dust ratio shows significantly lower dust masses at low metallicity / high redshift and higher dust temperatures.
The shape of the \Mdust evolution curves also changes in detail, likely a result of accretion or merger events bringing in preferentially metal-poor gas, where the reduced dust mass is further amplified by the variable dust-to-metals prescription.
}
\label{fig:app:RRsnap}
\end{figure}

In Figure~\ref{fig:app:RRsnap} we compare the impact of this metallicity-dependent prescription with our fiducial choice of the dust mass as a fixed fraction of the metal mass.
The effect is most pronounced at high redshift and low metallicity, where the metallicity-dependent gas-to-dust ratio (GDR) results in lower metal masses and higher mass-weighted dust temperatures, compared to the fiducial simulations with constant dust-to-metals ratios.

The details of the dust mass growth also change when adopting this prescription.
Accretion of low-metallicity gas brings in significantly less dust due to this second-order effect.
Relative to the fixed dust-to-metals model, the dust temperatures increase by up to $\sim50\%$ and the dust masses may drop by up two dex.
In sum, \Le typically decreases when considering a variable dust-to-metals situation.

\subsection{Impact on \HH Formation}

Efficient \HH formation relies on dust as a catalyst \citep[e.g.,][]{Gould1963}, so in addition to the reduction in dust mass, a variable dust-to-metals ratio should also affect the formation of \HH.
This results in an inconsistency in our treatment of the variable dust-to-metals as a reduced \HH formation efficiency may bring the snapshots in Figure~\ref{fig:RRdetect} back towards the empirical relation found for metal-rich galaxies.
The increased dust temperatures resulting from a reduced dust mass may also affect the \HH formation \citep[e.g.,][]{Cazaux2004}.
Assessing the magnitude of this effect would require a new suite of cosmological zoom simulations modifying the \HH formation model to explicitly consider the dust mass (rather than using the metallicity and assuming a constant dust-to-metals ratio) and is beyond the scope of this paper.
Future work with explicit treatment of dust formation and destruction (Q. Li et al. \emph{in preparation}) will be able to specifically address this inconsistency in our treatment.

However we emphasize that the low mass / low metallicity halos are the ones affected by this inconsistency.
Their lowered dust masses in the variable dust-to-metals scenario make them, at best, challenging to detect with current facilities.

\section{Sub-resolution Birth Clouds}
\label{app:birthcloud}

Dust reprocessing of the UV/optical radiation field can occur on the scales of the individual clouds in which star clusters form.
These ``birth clouds'' are below the resolution limit of our simulations, so including them requires the addition of a sub-grid model for the radiative transfer and time evolution.
To explore the impact of including birth clouds, we re-ran the radiative transfer for halo 401, using the \citet{Charlot2000} model as implemented in {\sc fsps} \citep{conroy09b} with the default values.

\begin{figure}
\centering
\includegraphics[width=0.45\textwidth]{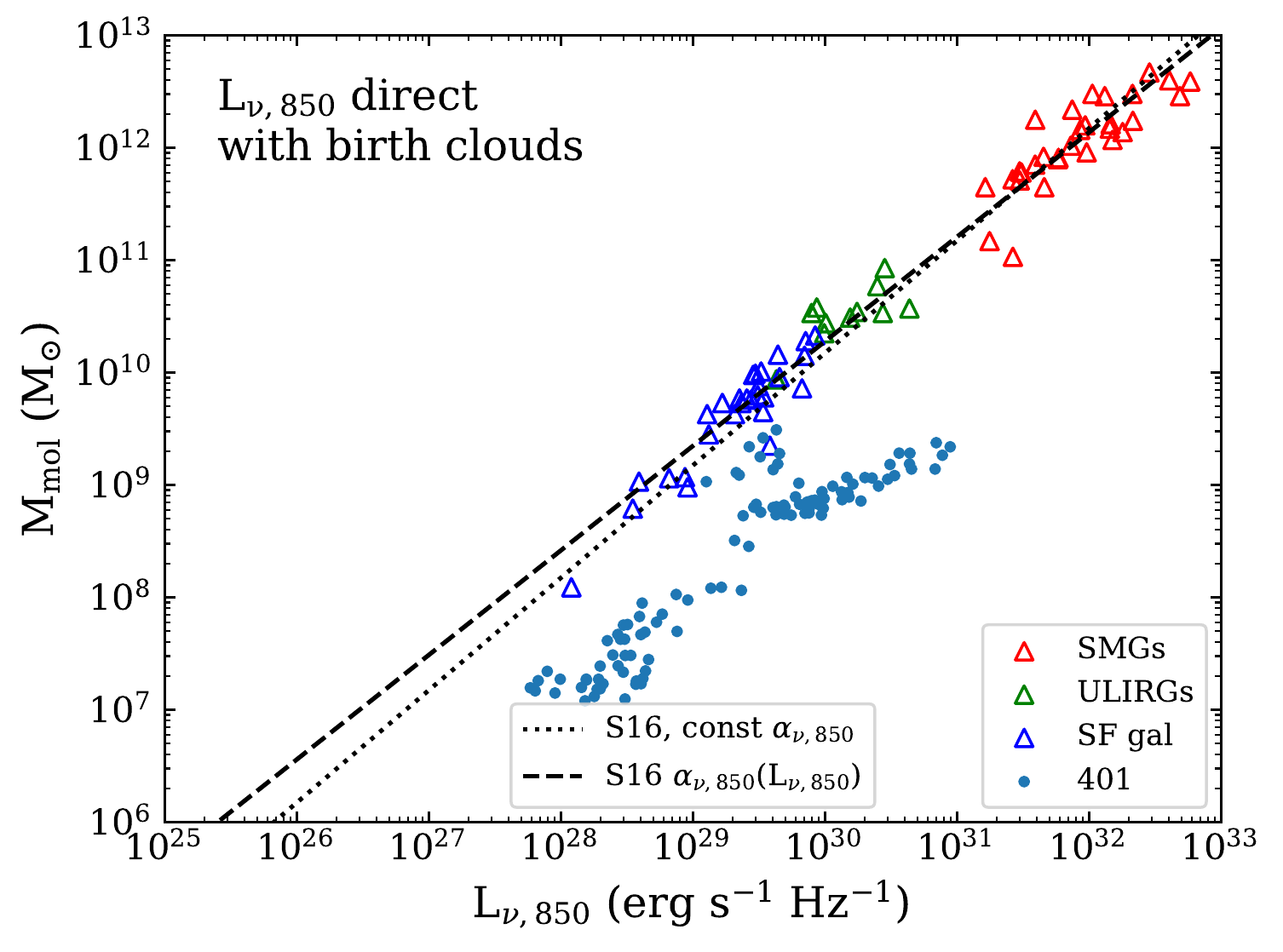}
\caption{The same as Figure~\ref{fig:S2016comp} but showing only halo 401 with the default birth cloud model of \citet{Charlot2000} as implemented in \citet{conroy09b}.
The \Le is significantly over-predicted at fixed \Mmol, compared to the observed galaxies presented in \citet{Scoville2016}.}
\label{fig:CFrelation}
\end{figure}

In Figure~\ref{fig:CFrelation} we reproduce Figure~\ref{fig:S2016comp} for the halo 401 snapshots with the birth cloud model enabled.
The inclusion of the birth cloud model still shows a relatively tight sequence, however it is offset to higher \Le by 1--2 dex.
This is likely due to compact dust in/around star forming regions resulting in a larger optical depth to IR photons and producing an overall cooler SED with significantly more emission at long wavelengths.
This appears inconsistent with the observations as it predicts an excess of \Le over what is observed.

As implemented in {\sc fsps+powderday} the birth cloud model likely produces too much opacity, as the dust within the birth clouds is not consistently modeled with the rest of the ISM in the hydrodynamic simulation.
This birth cloud model thus effectively means the ISM is too dust rich.
It may be possible to adjust the parameters of the birth cloud model to reduce the \Le excess.
However, the parameters for the birth clouds are largely unknown.
The default values from \citet{Charlot2000} were chosen to be consistent with galaxy-integrated observational properties of starbursts and as such may not be precise enough for the treatment of individual clouds in a sub-grid model.
There is ongoing debate regarding the timescale on which gas is cleared from the clusters \citep[see e.g.,][]{Prescott2007,Whitmore2014,Calzetti2015,Johnson2015,Murphy2018a}
Furthermore, this clearing timescale may be a function of the star cluster environment.

However, the fact that our simulations (Figure~\ref{fig:S2016comp}; without the birth cloud model) are consistent with the observations, this suggests that the radiative transfer effects and dust heating dominating the \Le emission of galaxies may occur on the scales which are robustly probed by our simulations (i.e., $\gtrsim10-50$~pc).

\end{document}